\documentclass[12pt]{iopart}
\usepackage{graphicx}
\usepackage{dcolumn}
\usepackage{epsfig}
\usepackage{color}
\usepackage{epstopdf}

\def\vtr#1{{\bf #1}}
\def\mtx#1{{\bf #1}}
\def\be{\begin{equation}}
\def\ee{\end{equation}}
\def\bea{\begin{eqnarray}}
\def\eea{\end{eqnarray}}
\def\ii{1\!\!1}

\def\modif#1{{ #1}}

\begin{document}
\article{~ \hfill\small{\rm preprint: NSF-KITP-07-01}}{}
\title{On the Sensitivity of a Hollow Sphere as a Multi-modal Resonant Gravitational Wave Detector}
%\title{Sensitivity of a Spherical Resonant Gravitational Wave Detector}
\author{F. Dubath$^1$, J. Extermann$^2$ and L. Gottardi$^3$}

\address{ $^1$ KITP,
Kohn Hall, UCSB
CA 93106, USA \footnote{On leave from DPT, Universit\'e de Gen\`eve, Switzerland}}
\address{$^2$  GAP - Biophotonics, Universit\'e de Gen\`eve,
20 rue de l'Ecole de M\'edecine, CH-1211 Gen\`eve 4, Switzerland}
\address{$^3$SRON National Institute for Space Research,
Sorbonnelaan 2,
3584 CA Utrecht, Netherlands}
\ead{dubath@kitp.ucsb.edu}
\begin{abstract}
We present a numerical analysis to simulate the response of a
spherical resonant gravitational wave detector and to compute its
sensitivity. Under the assumption of optimal filtering, we work out
the sensitivity curve for a sphere first taking into account only a
single transducer, and then using a coherent
analysis of the whole set of transducers. \\

We use our model for computing the sensitivity and therefore compare
different designs of spherical detectors. In particular we present
the case of 1 meter radius bulk and hollow spheres equipped with
transducers in TIGA configuration, and we explore the sensitivity of
a hollow sphere as a multi-modal
detector.
\end{abstract}

\submitto{\CQG}

\maketitle

\section{Introduction}
Direct detection of gravitational waves (GW) is still an unachieved
goal. The sensitivity improvement of single
detectors~\cite{DETECTORS},  to achieve the first detection, goes in
parallel with the set-up  of detector networks to perform real GW
astronomy~\cite{NETWORK}. To that purpose, future GW detectors
should be able to track GW arrival directions. Due to their
symmetry, spherical resonant detectors are natural GW `telescope'
candidates. Moreover, such detectors are based on the same
technology as resonant bars and can benefit of the decades of
experience gained by the GW community with this kind of experimental
setup.

A technical problem resides in the fact that big cryogenic resonant
spheres with high quality factor are difficult to build and to cool
down. One can partially bypass this problem using hollow spheres and
we will eventually focus on this kind of resonator.

However, the main limitation when dealing with resonant detectors,
is the small bandwidth compared with laser
interferometers~\cite{LIGO}.

Great progress have been made using the tuning of the electrical
mode~\cite{Auriga} which allows to enlarge the bandwidth from few Hz
to the order of 100 Hz around the resonance frequency. A possible
way to further enlarge the bandwidth consists of monitoring other
resonance frequencies of the antenna. In the case of a cylindrical
detector this strategy is not convenient since the coupling of the
$n$-th longitudinal mode to GW falls off as $(2n+1)^{-2}$. However
for a sphere the next mode quintuplet is tightly coupled to GW.
Monitoring these modes is possible using many transducers, tuned at
different frequencies. Here we explore this possibility and also
consider a model of a double-mode transducer sensitive to two modes
at different frequencies. We have to stress that both issues are
experimental challenges but they can lead to great improvement for
spherical detectors.

It is possible to enlarge the detector bandwidth by monitoring also
the second quadrupolar mode. In this way one could for example study
the radiation emitted by a binary system, consisting of either
neutron stars or black holes, in the in-spiral phase and determines
the chirp mass by measuring the time delay between excitations of
the first and second quadrupole modes of an hollow
sphere~\cite{Coccia:1996mq}. Used as a multi-modal detector even a
{\it single} sphere is able to set limit on the stochastic GW
background \cite{Lobo:2002it}. Furthermore, with an appropriate
resizing of the sphere, the second quadrupolar modes can be shifted
to the frequency region where existing small spherical detectors are
sensitive. This would open the possibility of coincidence search
between several spherical detectors and the DUAL detector
\cite{Bonaldi06}, building in this way  the base for a powerful
omnidirectional gravitational wave observatory.

In order to be able to compare different GW detectors we need to
work out the sensitivity with a ``standard'' approach and we decided
to use the strain sensitivity. A first goal of this paper is to
furnish a mathematical framework to describe all the internal noises
of a given detector and its response to an excitation (as example to
a GW burst). This is realized as a matrix model in the Fourier
space. Using this model, we set up numerical simulations and first
compute the sensitivity of each single transducer attached to the
sphere. We are fully aware that performing the data analysis in this
way does not exploit all the capacity of the spherical geometry.
However, this first computation is instructive and may be useful in
order to calibrate a future experiment. We point out the fact that
the noise coming from the other transducers will then be an
important limitation to the sensitivity. A clever way to use a
spherical detector is to perform the data analysis combining the
signals from all the transducers. Such a strategy, which requires
multidimensional matched filtering~\cite{Stevenson,FA,GottardiPRD}, exploits all
the geometrical properties of the detector and therefore gives best
sensitivity. This is the sensitivity to be compared with other
existing computations~\cite{sphere_sensibility}.

The second part of this paper will be dedicated to the application
of our model to ``realistic'' detectors having in mind the
comparison of different possible spherical resonant GW detectors. We
will focus on hollow spheres and in particular we explore the
different way to realize a multi-modal detector. The possibilities
we test are all technical challenges (dealing with 12 transducers,
transducer inside the hollow sphere, double mode transducers) but we
eventually show that there is no gain in working with transducer
inside the sphere and that 6 double mode transducers may compete
with 12 single transducers configurations only if they have a very
peculiar design.

\section{First part: the model}
\subsection{Description of spherical resonant GW detectors\label{DSRGW}}
\subsubsection{The  modes of a sphere}
The vibrational motion of a rigid body can be split into
eigen-modes. For a sphere there are two families of vibrational
eigen-modes: toroidal and spheroidal \cite{MichelsonZou}. Each mode
can be described as a forced damped harmonic oscillator,which means
that the $j$-th mode~\footnote{We use $j$ to describe collectively
all the numbers needed to specify the mode. For spheroidal modes
j=$\{n,l,m\}$~\cite{Merkowity95}. In the following we will work with
a subset of the modes. In particular we will use only $J$ spheroidal
modes that we will label by $j=1,..,J$} amplitude $z_j$ satisfies in
Fourier space the equation (see for example~\cite{Merkowity95}) \be
\label{modes}
\left(\omega_{s,j}^2-\omega^2+i\frac{\omega\omega_{s,j}}{Q_{s,j}}
\right)\tilde{z}_j(\omega)=\frac{1}{m_s}\tilde{F}_j(\omega)\ , \ee
where $m_s$ is the physical mass of the sphere, $\omega_{s,j}$ is
the $j$-th mode eigen-frequency, $Q_{s,j}$ its quality factor and
$\tilde{F}_j(\omega)$ the Fourier component of the forcing. The
forcing of the mode is due to a stochastic (Langevin) force and to
external forces. The stochastic force corresponds to the thermal
excitation of the sphere mode. Among the external forces we retain
only the tidal forces induced by GW, and forces due to the coupling
of the mode with the transducers.

We use  the frequencies $\omega_{s,j}$ as an input for our model.
The computation of the $\omega_{s,j}$ can be found in
\cite{Lobo:1995sc}.

\subsubsection{Transducer\label{SQUID}}
\begin{figure}
 \begin{center}
\includegraphics[width=8cm]{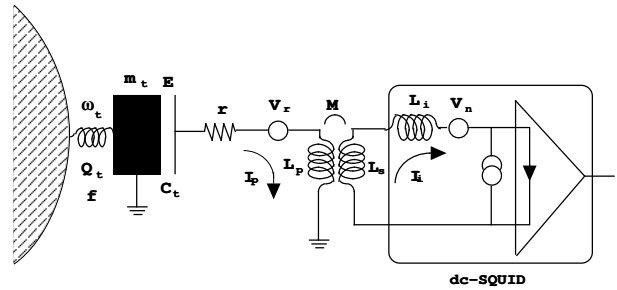} %[width=10cm]
\caption{\footnotesize The transducer model. %We show only the variables we use.
 \label{fig n1}}
 \end{center}
 \end{figure}

To give a clear presentation of the noises and the sensitivity we
need a full model for the transducer. We use capacitive transducers
(Figure~\ref{fig n1}), based on the one employed on bar
detector\cite{DETECTORS} and the spherical detector
MiniGRAIL~\cite{Gottardi}, composed by one resonator coupled to a
dc-SQUID with input transformer, where the SQUID is described as a
linear current amplifier~\cite{TescheClarke}. However our formalism
can be adapted for other kinds of transducer.

Each transducer (a mechanical resonator with its read-out) will be
modelled by a set of $p=3$ coupled differential equations, driven by
the specific intrinsic noises and its coupling to the sphere. In
Fourier space each transducer and each read-out are described by the
equation set \be\label{transducer}
\pmatrix{m_{t}\left(\omega_{t}^2-\omega^2+i\frac{\omega\omega_{t}}{Q_{t}}
\right) & -iE/\omega &0 \cr E&\!\!\!\!\!\!\! \!\!\!\! \!r+i(\omega
L_{p}-\frac{1}{(\omega C_{t})})&\!\!\!\!-iM\omega\cr 0 & -iM\omega
&\!\!\!\!i \omega(L_{s}+L_{i})}
\pmatrix{\tilde{x}\cr\tilde{I}_{p}\cr\tilde{I}_{i}}
=\pmatrix{\tilde{f}\cr\tilde{V}_{r}\cr\tilde{V}_{n}} \ee where
$m_{t},\ \omega_{t},\ Q_{t}$ are the resonators mass,
eigen-frequency and quality factor, $E$ is the electric field in the
capacitor $C_t$ formed with the resonator, $r$ the transformer
resistance, $L_{i},\ L_{p},\ L_{s}$  the inductances of the SQUID
input coil, primary and secondary of the transformer,
 and $M$ the mutual inductance of the transformer; see Figure~\ref{fig n1}.
$\tilde{x}$ is the transducers position~\cite{SphereAlice},
 $\tilde{I}_{p},\ \tilde{I}_{i}$ are
the currents into the transformer and the SQUID, and
$\tilde{I}_{i}$ is the measured quantity.
Finally, $\tilde{V}_{r}$ and $ \tilde{V}_{n}$ are the voltages corresponding to the intrinsic noise
and $\tilde{f}$ is the sum of the thermal force and the forces due to the coupling to the
spheres modes, which are described in Section~\ref{S coupling}.\\

In what follow, we will use a set of $N$ transducers and therefore
add an index $k=1,..,N$ to all the quantities describing the
transducer.

\subsubsection{Double-mode transducer\label{SQUID2}}

Having in mind the possibility of building a multi-modal spherical
resonant GW antenna we note that a simple way to achieve this goal
is the use of multi-modal transducers. A multi-modal transducer has
to be designed in order to be resonantly coupled to two distinct
frequencies. As a model, one can think to a multi-modal resonator
with different vibration frequencies, (each frequency tuned on one
of the sphere resonance frequencies) and corresponding electronic
read-out. The resonator can be modelled as two oscillators located
at the same point of the sphere.

\begin{figure}
 \begin{center}
\includegraphics[width=6cm]{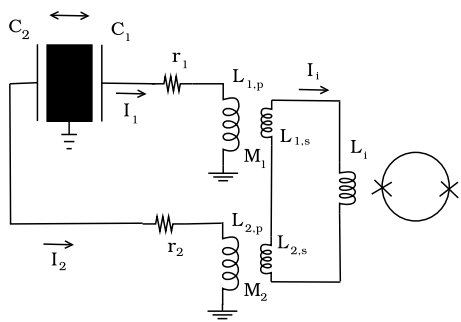} %[width=10cm]
\caption{\footnotesize The double transducer model. %We show only the variables we use.
\label{fig nd}}
 \end{center}
 \end{figure}

In Figure~\ref{fig nd}, we show the schematic diagram of  a
capacitive double mode transducer composed of a resonator
mechanically coupled to the two first spheroidal modes of the sphere
and a single SQUID read-out. The two resonating mass schematically
drawn in the figure should be considered fixed at the same point on
the sphere surface. Such a transducer could be fabricated using, for
example, a mushroom type resonator placed inside a ring-shaped
membrane transducer  or using the geometry suggested in \cite{Bassan97} . By means of two electrically insulated
electrodes and two super-conducting  matching transformers, one is
able to couple the signal from both the resonators to a single SQUID
amplifier, simplifying greatly the detector read-out electronics and
the cryostat wiring.To achieve an optimal impedance matching each
electrical mode defined by the two LC circuits should be coupled to the two corresponding
spheroidal modes. In this scheme a mixing may occur between the two
modes. However we will show here that  this will not significantly
reduce the detector sensitivity.

The double-mode transducer  is described by a set of $p=5$
equations: two for the mechanical modes, two for the transformer
currents $I_1,I_2$, and one for the SQUID input current $I_i$:
\begin{eqnarray}
&&\left(\omega_{t1}^2-\omega^2+i\frac{\omega\omega_{t1} }{Q_{t1} } \right) \tilde{x}_{t1} -i\frac{E_1}{\omega}\tilde{I}_1=\tilde{f}_1\\
&&\left(\omega_{t2}^2-\omega^2+i\frac{\omega\omega_{t2} }{Q_{t2} } \right) \tilde{x}_{t2} -i\frac{E_2}{\omega}\tilde{I}_2=\tilde{f}_2\\
&& E_1\tilde{x}_{t1} +(r_1+i\omega L_{1,p}-\frac{i}{\omega C_1})\tilde{I}_1 -i\omega M_1\tilde{I}_i=\tilde{V}_{1,r}\\
&&E_2\tilde{x}_{t2}+(r_2+i\omega L_{2,p}-\frac{i}{\omega C_2})\tilde{I}_2 -i\omega M_2\tilde{I}_i=\tilde{V}_{2,r}\\
&& i\omega( L_{1,s}+L_{2,s}+L_{i})\tilde{I}_i-i\omega (M_1\tilde{I}_1
+ M_1\tilde{I}_2)=\tilde{V}_n{\tt ,}
\end{eqnarray}
where $E_i,\ i=1,2$ is the electrical field in the $i$-th capacitors
and $\tilde{x}_{ti}$ the displacement of the two mechanical modes.
\bea\label{hollowtransducer}
\hspace{-3cm}\pmatrix{\left(\omega_{t1}^2-\omega^2+i\frac{\omega\omega_{t1}}{Q_{t1}}
\right) & 0 & -i\frac{E_1}{\omega} & 0 & 0\cr 0 &
\left(\omega_{t2}^2-\omega^2+i\frac{\omega\omega_{t2}}{Q_{t2}}
\right) & 0 & -i\frac{E_2}{\omega} & 0 \cr E_1 & 0 &\!\!\!\!\!\!\!
\!\!\!\! \!r_1+i(\omega L_{1,p}-\frac{1}{(\omega C_{1})})& 0
&\!\!\!\!-iM_1\omega\cr 0 & E_2 &\!\!\!\!\!\!\! \!\!\!\!
\!r_2+i(\omega L_{2,p}-\frac{1}{(\omega C_{2})})& 0
&\!\!\!\!-iM_2\omega\cr 0 & 0 & -iM_1\omega & -iM_2\omega &\!\!\!\!i
\omega(L_{1,s}+L_{2,s}+L_{i})}\times\nonumber \\
\pmatrix{\tilde{x}_{t1}\cr\tilde{x}_{t2}\cr\tilde{I}_{1}\cr\tilde{I}_{2}\cr\tilde{I}_{i}}
=\pmatrix{\tilde{f}_{1}\cr\tilde{f}_{2}\cr\tilde{V}_{1,r}\cr\tilde{V}_{2,r}\cr\tilde{V}_{n}}
\eea

\subsubsection{Coupling transducers to the modes of the sphere\label{S coupling}}

Since none of the state-of-the-art transducers are able to couple
efficiently to tangential motion, we will restrict ourself to
transducers only sensitive to radial motion of the sphere, and
consequently we will neglect the coupling to toroidal
modes~\cite{MichelsonZou}.

We call $(\theta_k, \phi_k)$ the location of the $k$-th transducer.
At this position on the sphere surface, the $j=\{n,l,m\}$-th
spheroidal mode induces a radial displacement described

\be \label{alpha}
\alpha_{nl}(R_s)Y_{lm}(\theta_k,\phi_k)\equiv\alpha_jB_{jk} \ee

where $\alpha_j\equiv\alpha_{nl}(R_s)$ is the radial eigen-function
evaluated at the sphere surface, and $Y_{lm}$ is a spherical
harmonic. This equation defines the pattern matrix $B_{jk}$. Using
this matrix we obtain the equation of motion
(~\cite{Merkowity95},~\cite{GottardiPRD}). The coupling between the
sphere modes and the transducers is then given (in the Fourier
space) by \bea
\tilde{F}_j&=&\tilde{F}^{\rm noise+GW}_j+\sum_k \alpha_jB_{jk}\left(\left(\omega_{t,k}^2+i\frac{\omega\omega_{t,k}}{Q_{t,k}} \right)m_{t,k}\tilde{x}_k-\tilde{f}^{\rm noise}_k\right)\label{couplage1}\\
\tilde{f}_k&=&\tilde{f}^{\rm noise}_k+ m_{t,k}\omega^2\sum_jB_{jk}\alpha_j\tilde{z}_j\ .\label{couplage2}
\eea

We obtain a set of ($J +p N$) coupled equations, with $J$ the number
of modes taken into account that we discuss bellow, $N$ the number
of transducers and $p$ the number of equations need to describe a
single transducer.

It is important to note that generally a transducer is coupled with
a subset of the five sphere modes and therefore the presence of
transducers provides an indirect coupling between the different
modes. Reciprocally, different transducers are coupled through the
sphere modes. Therefore, the total noise spectrum of a given
transducer has a contribution from the intrinsic noises of the other
transducers.

We are principally interested in the quadrupolar ($\ell =2 $)
spheroidal mode family. The lowest quadrupolar multiplet
($n=1,\ell=2$) has the advantage to contain the 5 spheroidal modes
with the lowest resonance frequency~\cite{Lobo:1995sc}. When we are
only interested in those modes we can reduce the sphere vibration to
these 5 modes and neglect the effect of the modes at higher
frequency. When we describe a multi-modal detector, we are also
interested in the second quadrupolar multiplet ($n=2,\ell=2$) at about
2 times the fundamental mode resonance. We note
that other modes, not coupled to GW, ($n=1,\ell=0,1,3,4$) have their resonance frequency
between the ones of the first and second quadrupolar
multiplet~\cite{Lobo:1995sc}. These modes
induce additional thermal noise sources and have to be included when
calculating the sensitivity of a multi-modal resonator, like previously shown in \cite{Briant}.

\subsubsection{Bulk sphere model}
The remaining of the computation is easier to follow if applied to a
concrete example. For this purpose, we present a simple model with
only the spheroidal quadrupole modes of the sphere ($J=5,\ l=2$),
and with $N=6$ capacitive transducers placed in the TIGA
configuration~\footnote{The precise location of the transducer
changes the $B$ matrix causing to loose the TIGA configuration.
However this do not change the following analysis}
~\cite{Merkowity95}. Note that we can write the system of ($J +p N$)
equations in the same form independently of the number of modes and
transducers. We also use this example in order to set the notations.

Collecting the equations
(\ref{modes},\ref{transducer},\ref{alpha},\ref{couplage1}) and
(\ref{couplage2}) we obtain a description of the entire detector as

\be\label{eom}
\underbrace{\pmatrix{\setlength{\unitlength}{1 pt}   %% set to 1pt for draft version and 0.7 pt for the 2 column version
 \begin{picture}(88,100)\put(0,0){\line(0,1){88}}\put(22,0){\line(0,1){88}}\put(88,0){\line(0,1){88}}\multiput(22,22)(3,0){22}{\line(1,0){2}}\put(0,44){\line(1,0){22}}
 \multiput(22,44)(3,0){22}{\line(1,0){2}}
\put(0,0){\line(1,0){88}}\put(0,88){\line(1,0){88}}\put(0,66){\line(1,0){88}}
\multiput(44,0)(0,3){22}{\line(0,1){2}}\put(44,66){\line(0,1){22}}
\multiput(66,0)(0,3){22}{\line(0,1){2}}
\multiput(0,22)(3,0){2}{\line(1,0){2}}\multiput(22,22)(-3,0){2}{\line(-1,0){2}}
\multiput(66,66)(0,3){2}{\line(0,1){2}}\multiput(66,88)(0,-3){2}{\line(0,-1){2}}
\put(8,18){$0$}
\put(62,73){$0$}
\put(7,73){$\mtx{S}$}
\put(5,51){$\mtx{C}_2$}
\put(27,73){$\mtx{C}_1$}
\put(51,28){$\mtx{T}$}
 \end{picture}}}_{\mtx{Z}}\setlength{\unitlength}{1 pt} \pmatrix{\tilde{\vtr{z}}\cr\tilde{\vtr{q}}\cr\tilde{\vtr{I}}_{p}\cr\tilde{\vtr{I}}_{i}} =\underbrace{\pmatrix{\ii_5&-\alpha \mtx{B}& 0&0\cr0&\ii_6&0&0\cr0&0&\ii_6&0\cr0&0&0&\ii_6}}_{\mtx{A}}\pmatrix{\tilde{\vtr{F}}^{\rm noise+ GW}\cr\tilde{\vtr{f}}^{\rm noise}\cr\tilde{\vtr{V}}_{r}\cr\tilde{\vtr{V}}_{n}}
\ee
where $\tilde{\vtr{z}}=\pmatrix{\tilde{z}_1\cr\vdots\cr\tilde{z}_5}$,
$\tilde{\vtr{q}}=\pmatrix{\tilde{x}_1\cr\vdots\cr\tilde{x}_6}$, and so on for
the other variables.\\
$\mtx{S}$ is a $5\times 5$ diagonal sub-matrix given by the left
hand side (LHS) of (\ref{modes}) , $\mtx{T}$ is a
$3\cdot6\times3\cdot6$ sub-matrix with structure given by the LHS of
(\ref{transducer}) (each number in (\ref{transducer}) is now a
$6\times6$ diagonal matrix). The matrices $\mtx{C}_1$ and
$\mtx{C}_2$ are $5\times6$ (resp. $6\times5$) sub-matrix read out of
(\ref{couplage1},\ref{couplage2})~\footnote{As we have only modes
with $l=2$, all the $\alpha_j$ are equal. Therefore, we drop the
indices.}  which describe the mechanical coupling between the
spheroidal modes and the transducer modes and are given by \bea
\mtx{C}_1&=&-\alpha \mtx{B} {\rm Diag}\left(m_{t,k}(\omega_{t,k}^2+i\frac{\omega\omega_{t,k}}{Q_{t,k}})\right)\\
\mtx{C}_2&=&-\alpha\omega^2{\rm Diag}(m_{t,k}) \mtx{B}^T \ .
\eea

\subsection{Detector noise description\label{NOISE}}
Starting from equation (\ref{eom}), knowing the forces and the voltage acting
on the detector allows us to compute the sphere modes $\tilde{\vtr{z}}$,
the displacement $\tilde{\vtr{q}}$ of the transducer, and
the currents $\tilde{\vtr{I}}_p$ and $\tilde{\vtr{I}}_i$.
To do that we invert the $\mtx{Z}$  matrix and rewrite (\ref{eom}) as
\be\label{G}
\pmatrix{\tilde{\vtr{z}}\cr\tilde{\vtr{q}}\cr\tilde{\vtr{I}}_{p}\cr\tilde{\vtr{I}}_{i}}=\underbrace{\mtx{Z}^{-1}\mtx{A}}_{\mtx{G}}\pmatrix{\tilde{\vtr{F}}^{\rm noise+ GW}\cr\tilde{\vtr{f}}^{\rm noise}\cr\tilde{\vtr{V}}_{r}\cr\tilde{\vtr{V}}_{n}}
\ee
Mode amplitudes, transducer displacement and transformer
currents are not directly measured and we are therefore interested only in the $N$
last lines of $\mtx{G}$, which give the proportionality coefficients between
forces (and voltages) and SQUID input currents $\tilde{\vtr{I}}_i$.

We now describe the different noise contributions.
\subsubsection{Noise description}

We restrict ourself to the case where the disturbances $\vtr{F}^{\rm
noise},\ \vtr{f}^{\rm noise}$ and $\vtr{V}_{r}$ are only due to
thermal excitations. In this case we can only access statistical
property of these forces that is:
\begin{eqnarray}
\langle F(t)\rangle =0&\hspace{10mm} &\langle V(t)\rangle =0\ ,\\
\langle F(t)F(t')\rangle =A_0\delta(t-t')&\hspace{10mm} &\langle V(t)V(t')\rangle =A^e_0\delta(t-t')\ .
\end{eqnarray}
Furthermore one can compute~\cite{MM}  that the coefficient $A_0$ takes the form
\be
A_0=2k_BTm/Q\hspace{10mm} A_0^e=2k_BTr\ .
\ee
where $k_B$ is the Boltzmann constant, $T$ is the
thermodynamic temperature of the detector, and $r$ the transformer resistance.

Using the definition of the single-sided spectral density of the force which is obtained through the autocorrelation
\be
\langle F(t)F(t')\rangle=\frac{1}{2}\int_0^\infty\frac{d\omega}{2\pi}S_F(\omega)e^{-i\omega(t-t')}\ ,
\ee
we obtain
\bea
\vtr{S}_{F,j}&=&4m_s\omega_{s,j}\frac{k_BT}{Q_{s,j}}\label{nm1}\\
\vtr{S}_{f,k}&=&4m_{t,k}\omega_{t,k}\frac{k_BT}{Q_{t,k}}\label{nm2}\\
\vtr{S}_{V_r,k}&=&4k_BTr_k.\label{ne1}
\eea

\begin{figure}[h]
 \begin{center}
\includegraphics[width=6cm]{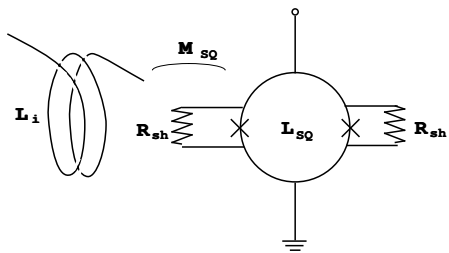}
\caption{\footnotesize SQUID detail.% We show only the variables we use.
 \label{fig n2}}
 \end{center}
 \end{figure}

The SQUID has also intrinsic noises which can be split into voltage
and current noises. The determination of these noises requires the
knowledge of the complete SQUID design. Therefore, we have to go
beyond the description of Section~\ref{SQUID}. In particular we need
the shunt resistance $R_{sh,k}$ used to remove hysteresis, the
washer inductance $L_{SQ,k}$, and $M_{SQ,k}$ the mutual inductance
between the SQUID input and the
 washer, see Figure~\ref{fig n2}. \\
\\
The SQUID voltage and current noises are, according to Clarke's model ~\cite{TescheClarke}
\bea
\vtr{S}_{V_n,k}&=&11\frac{k_BT_{SQ,k}}{R_{sh,k}}\omega^2M^2_{SQ,k}\label{ban}\\
\vtr{S}_{W,k}&=&16\left(\frac{L_{SQ,k}}{M_{SQ,k}}\right)^2\frac{k_BT_{SQ,k}}{R_{sh,k}}\label{lnoise}
\eea
where $T_{SQ,k}$ is the SQUID thermodynamic temperature.

\subsubsection{Noise matrix}
 The noise transducer outputs are proportional to the SQUID input current and thus given by
\be
 \tilde{\vtr{I}}_i(\omega)=\mtx{M}_i\mtx{G}(\omega)\pmatrix{\tilde{\vtr{F}}^{\rm noise}\cr\tilde{\vtr{f}}^{\rm noise}\cr\tilde{\vtr{V}}_r\cr\tilde{\vtr{V}}_n}(\omega)+\tilde{\vtr{I}}_W(\omega)\equiv\mtx{G}_I(\omega)\tilde\vtr{F}(\omega) +\tilde{\vtr{I}}_W(\omega)
 \ee
 where $\tilde{\vtr{I}}_W(\omega)$ is the SQUID noise current and $\mtx{M}_i$ is a $N\times(3N+J)$ matrix given by
 \be
 \mtx{M}_i=\pmatrix{0_{N,J} &0_{N,N}&0_{N,N}&\ii_{N}}
 \ee used in order to conserve only the SQUID input current and  we have defined $\mtx{G}_I(\omega)\equiv\mtx{M}_i\mtx{G}(\omega)$  to simplify the notation.\\

Knowing the current we can compute the noises matrix $\mtx{S}$ , which is given by
\be
\mtx{S}=\vtr{I_i}(\omega)\vtr{I_i}^\dagger(\omega)=\mtx{G}_I\tilde\vtr{F}\tilde\vtr{F}^\dagger\mtx{G}_I^\dagger+ \mtx{G}_I\tilde\vtr{F}\tilde{\vtr{I}}_W^\dagger+\tilde{\vtr{I}}_W\tilde\vtr{F}^\dagger\mtx{G}_I^\dagger+\tilde{\vtr{I}}_W\tilde{\vtr{I}}_W^\dagger \label{noisematrix}
\ee

The matrix  $\tilde\vtr{F}\tilde\mtx{F}^\dagger$ is the correlation matrix of the different forces and voltages (Note that not all entries have the same units). Assuming that the different forces and voltage are only due to noises and are not-correlated this reduce to a diagonal matrix containing only autocorrelation
\be
\vtr{F}\mtx{F}^\dagger={\rm Diag}\left(\vtr{S_F},\vtr{S_f},\vtr{S_{V_r}},\vtr{S_{V_n}}\right)
\ee
The same argument leads to the cancellation of the $\tilde\vtr{F}\tilde\mtx{I}_W^\dagger$ and $\tilde\vtr{I_W}\tilde\mtx{F}^\dagger$ terms into (\ref{noisematrix}) and to set
\be
\tilde{\vtr{I}}_W\tilde{\vtr{I}}_W^\dagger={\rm Diag}(\vtr{S}_W)
\ee

Therefore under the assumption that the different noises are not correlated the noise matrix (\ref{noisematrix}) reduces to
\be
\mtx{S}^{\rm noise}=\mtx{G}_I{\rm Diag}\left(\vtr{S_F},\vtr{S_f},\vtr{S_{V_r}},\vtr{S_{V_n}}\right)\mtx{G}_I^\dagger+{\rm Diag}(\vtr{S}_W)\label{noisematrix2}
\ee

\subsection{Effect of a GW \label{GWE}}
The presence of a GW will manifest itself as a force acting on the
sphere modes. We skip the computation (see ~\cite{Merkowity95}) and
just note that in Fourier space the force acting on the $j$-th mode
is \be \label{chi} \tilde{F}^{\rm GW}_j=-\frac{1}{2}\omega^2
m_s\chi_j R_s \tilde{h}_j \ee where $\chi_jR_s$, the effective
length of the mode, depends only on the multiplet ($n$) to which the
mode belongs, and $\tilde{h}_j$ is the projection of the (spatial
part of the) GW tensor~\footnote{By a GW tensor we mean the
perturbation of the background metric in the TT gauge} on the $j$-th
mode. Choosing the decomposition of the tensor on real
matrix~\cite{SphereAlice} we specify the form of $\tilde{\vtr{h}}$
for the quadrupolar modes $(\ell=2)$ by\footnote{Note that if we
describe gravitation by general relativity, only  spheroidal
quadrupolar modes ($l=2$) have non zero $\tilde{h}_j$}. \be
\tilde{\vtr{h}}=\mtx{T}_V\pmatrix{\tilde{h}_+\cr\tilde{h}_\times}\equiv\pmatrix{\frac{\sqrt{3}}{2}\sin^2
\theta &0 \cr -\frac{1}{2}\sin 2\theta\sin\phi&\sin\theta\cos\phi\cr
\frac{1}{2}\sin 2\theta\cos\phi&\sin\theta\sin\phi       \cr
\frac{1}{2}\left(1-\cos^2\theta\right)\cos2
\phi&\cos\theta\sin2\phi\cr
-\frac{1}{2}\left(1-\cos^2\theta\right)\sin2
\phi&\cos\theta\cos2\phi}\pmatrix{\cos 2\psi&\sin 2 \psi\cr -\sin 2
\psi &\cos 2 \psi}\pmatrix{\tilde{h}_+\cr\tilde{h}_\times} \ee where
$(\theta,\phi)$ gives the arrival direction.  $\tilde{h}_+$ and
$\tilde{h}_\times$ describe the two polarization of the GW in the
source frame. The actual polarization in the detector depend on the
position of the source $(\theta,\phi)$ and on the orientation of the
source which is rotated by an angle $\psi$ along the line of the
sight with respect to the detector frame. We call $\psi$ the
polarization angle since for a source at a given location a change
in $\psi$ act only on the polarization.

The presence of a GW in a noise-free detector leads to an output
signal given by the vector of SQUID input current \be
 \vtr{I}_{\rm sig}(\omega)=\mtx{G}_I(\omega)\pmatrix{\tilde{\vtr{F}}^{\rm GW}(\omega)\cr\vtr{0}\cr\vtr{0}\cr\vtr{0}} =\hat{\mtx{G}}_I(\omega)\tilde{\vtr{F}}^{\rm GW}(\omega)
=-\frac{1}{2}\omega^2 m_sR_s\chi\hat{\mtx{G}}_I(\omega) \mtx{T}_V\pmatrix{\tilde{h}_+\cr\tilde{h}_\times}\label{Isig}
\ee
where we have assumed that all $\chi_j$ have the same value $\chi$ and we have defined $\hat{\mtx{G}}_I$, the sub-matrix
\be
\hat{\mtx{G}}_I=\mtx{G}_I\pmatrix{\ii_J\cr0_{N,J}\cr0_{N,J}\cr0_{N,J}}=\underbrace{\pmatrix{0_{N,J} &0_{N,N}&0_{N,N}&\ii_{N}}}_{N\times(3N+J)}\mtx{G}\underbrace{\pmatrix{\ii_J\cr0_{N,J}\cr0_{N,J}\cr0_{N,J}}}_{(3N+J)\times J}
\ee

\subsection{Detector sensitivity}

The detector sensitivity is given by the comparison between the
output due to the noise and the one due to a GW. The output of the
detector is given by $N$ output currents (one by transducer). One
can combine theses outputs in different way. The ability of
extracting the GW signal depends on the way the outputs are combined
and so do the sensitivity.

\subsubsection{Single transducer analysis}

In this first data analysis schema, we considered a rather naive way
of using the different transducers: each one is taken as an
independent experiment. This has the advantage of simplifying
drastically the data analysis (which can be based on the one of the
resonant bar without further modifications). However it is clear
that in this way we are loosing most  of the advantages of the
spherical geometry. Nevertheless it is instructing to perform this
analysis having in mind that it will probably be useful in the
first phase of any sphere experiment.\\

If we are looking only at the $k$-th transducer, its noise spectral
density is given by the diagonal components of (\ref{noisematrix2})
\bea
S_{I_i^k}&=&\mtx{S}^{kk}=\sum_{\ell=1}^j\vert\mtx{G}_{J+2N+k,\ell}\vert^2\vtr{S}_{F,\ell}+\sum_{\ell=1}^N\vert\mtx{G}_{J+2N+k,J+\ell}\vert^2\vtr{S}_{f,\ell}\nonumber  \\
&&+\sum_{\ell=1}^N\vert\mtx{G}_{J+2N+k,J+N+\ell}\vert^2\vtr{S}_{V_n,\ell}+\sum_{\ell=1}^N\vert\mtx{G}_{J+2N+k,J+2N+\ell}\vert^2\vtr{S}_{V_r,\ell}
+ \vtr{S}_W^k \label{noisesingle} \eea

and this has to be compared with the spectral density of the $k$-th
SQUID input current due to the GW.  The latter is obtained as the
square of the equ. (\ref{Isig})

\bea
S^{GW}_k&=&\frac{1}{4}\omega^4m_s^2R_s^2\chi^2\left\vert\sum_{\ell
m} \mtx{G}_I^{k\ell}\mtx{T}_V^{\ell
m}\pmatrix{\tilde{h}_+\cr\tilde{h}_\times}^m\right\vert ^2 \eea

If we know the GW arrival direction, we can compute each $\mtx{T}_V$
and simulate the detector response. The inverse problem, computing
the GW propagation direction, needs high signal to noise ratio (SNR)
~\cite{Stevenson}. If we are interested in the detector sensitivity
we are working at SNR=1, and therefore we have no information on the
arrival direction. Furthermore, if we choose an arbitrary direction
we can be in a case where some of the modes are poorly or not at all
coupled to this peculiar GW. Consequently we can underestimate the
sensitivity~\footnote{As an illustration: the case of a cylindrical
detector. If we choose the GW arrival direction parallel to the
detector axis, the GW is not seen and no information is given about
the detector sensitivity.}. As the spectral density is a statistical
feature of the signal, it is then natural to perform an average on
the possible arrival directions and over the source orientation
(polarization angle).

Averaging $S^{GW}_k$ over the arrival direction ($\theta,\phi$) and
the source orientation (polarization angle $\psi$) and assuming an
intrinsic polarization $\tilde{h}_+\equiv\tilde{h},\
\tilde{h}_\times\equiv0$, we find ~\cite{Extermann} \bea
S^{GW}_k&=&\frac{1}{10}\omega^4m_s^2R_s^2\chi^2\tilde{h}^2\sum_{\ell=1}^5\mtx{G}^2_{J+2N+k,\ell}.
\eea

The strain sensitivity is given by the comparison $S_{I_i^k}=S^{GW}_k$ and is
\be
\tilde{h}_c(\omega)=\left(\frac{S_{I_i^k}}{\frac{1}{20}\omega^4m_s^2R_s^2\chi^2\sum_{\ell=1}^5\mtx{G}^2_{J+2N+k,\ell}}\right)^{1/2}\equiv\left(\frac{S_{I_i^k}(\omega)}{T_F(\omega)}\right)^{1/2}, \label{Tf1}
\ee
where we have defined the transfer function $T_F$.

\subsubsection{Coherent analysis}

Rather than looking at each transducer separately we can perform a
coherent analysis taking into account all the $N$ signals. In this
case we have to perform multidimensional matched filtering
\cite{Stevenson}. In the special case of TIGA configuration, one can
use the mode channels in order to obtain directly the motion of the
quadrupolar modes of the sphere \cite{Merkowity95}. However, on one
hand, this property was proved for the sphere with mechanical
resonators and may be affected by the tuned electric oscillators
and/or by the perturbations induced by the suspension of the sphere
in the Earth gravity. On the other hand, having the motion of the
sphere modes is not enough: one has to make the deconvolution
between the excitations and the damped motion of the modes. \\
If, for a given transducer number and configuration, we know the detector
response to an excitation, one can directly build a multidimensional matched
filter \cite{FA}. One can therefore achieve optimal filtering (once the waveform
of the GW is know or assumed to be know). Note that the construction of the
filter can be performed using our numerical model, see  \cite{FA} for details.\\
The multidimensional matched filtering can be defined and used for
{\it any} configuration but in general it has to be computed
numerically. However the TIGA configuration offers the great
advantage of accepting analytic solutions to the deconvolution
problem. In the following, thanks to a result of
Stevenson~\cite{Stevenson}, we do not need to know how the matched
filtering will be performed and whether the filter is obtained in a
numerical or analytical way. We only need to know that an optimal
filter is applied to the data. In this case the SNR is given by
\cite{Stevenson} \bea
SNR&=&\int_{-\infty}^{\infty}\sigma(\omega)d\omega/2\pi\\
\sigma(\omega)&=&\vtr{I}_{\rm sig}^\dagger (\omega)\mtx{S}^{-1}(\omega)\vtr{I}_{\rm sig}(\omega)
\eea
where $\mtx{S}$ is the noise matrix defined by equation~(\ref{noisematrix}).\\
Using the expressions (\ref{Isig}), we get \be
\sigma(\omega)=\frac{1}{4}\omega^4m_s^2R_s^2\chi^2
\pmatrix{\tilde{h}_+^*&\tilde{h}_\times^*}\mtx{T}_V^\dagger\underbrace{
\hat\mtx{G}_I^\dagger\mtx{S}^{-1}\hat\mtx{G}_I}_{\mtx{H}}\mtx{T}_V\pmatrix{\tilde{h}_+\cr\tilde{h}_\times}
\ee

In order to evaluate this expression we face again the problem that
$\mtx{T}_V$ is a function of the arrival direction and of the source
orientation.  As in the single transducer analysis it is convenient
to average the expression over the direction and the polarization.
This gives

\bea
\langle\pmatrix{\tilde{h}_+\cr\tilde{h}_\times}^\dagger\mtx{T}_V^\dagger\mtx{H}\mtx{T}_V\pmatrix{\tilde{h}_+\cr\tilde{h}_\times}\rangle&=&\int_0^\pi\int_0^{2\pi}\frac{\sin(\theta)d\theta d\phi}{4\pi}\int_0^{\pi}\frac{d\psi}{\pi}\pmatrix{\tilde{h}_+\cr\tilde{h}_\times}^\dagger\mtx{T}_V^\dagger\mtx{H}\mtx{T}_V\pmatrix{\tilde{h}_+\cr\tilde{h}_\times}\nonumber\\
&=&\frac{1}{5}{\rm Tr}(\mtx{H})\left( \tilde{h}_+^2+\tilde{h}_\times^2\right)
\eea

assuming again a polarized GW  $\tilde{h}_+\equiv\tilde{h},\
\tilde{h}_\times\equiv0$ we get for $\langle\sigma\rangle$ \be
\langle\sigma\rangle=\frac{1}{20}\omega^4m_s^2R_s^2\chi^2\tilde{h}^2{\rm
Tr}(\mtx{H}) \ee

The strain sensitivity (averaged on the sky and the source
orientation) is then given by solving the equation
$\langle\sigma\rangle\equiv1$  for $\tilde{h}(\omega)$. We
eventually obtain \be
\tilde{h}_c(\omega)=\frac{2\sqrt{5}}{\omega^2m_SR_S\chi\sqrt{{\rm
Tr}(\mtx{H})}}\label{sensicoh} \ee

\section{Second part: application and simulations}
\subsection{The filled sphere\label{a}}

We start with the case of a filled sphere, this case is easier to
compare with existing computations~\cite{sphere_sensibility} and can
be used as a benchmark for the hollow sphere simulations. We also
use this simple case in order to highlight the how the single
transducer sensitivity is modified by the presence of other
transducers. With a numerical code, implementing
(\ref{eom}),(\ref{G}-\ref{lnoise}),(\ref{noisesingle}), (\ref{Tf1}),
and (\ref{sensicoh}), we can compute the strain sensitivity for a
spherical resonant mass with capacitive transducers coupled to a
dc-SQUID with input transformer and resonances frequency of the
transducer tuned in order to monitor the five quadrupolar modes. The
parameters used in our simulation can be found in the
Table~\ref{table 1}.

\begin{table}[htdp]

\begin{center}
\begin{tabular}{|c|c|c|}

\hline
&Radius& $R_s=1[m]$\\
&Mass&$m_s=33091[kg]$     \\
&Quality factor&$Q_s=5\cdot10^7$\\
Sphere&Number of modes&$  J=5$\\
&Modes frequencies&$\frac{\omega_{s,j}}{2\pi}=1049,1053,1057,1061,1065~[Hz]$\\
&Radial eigen-function&$\alpha=-2.89$\\
&Ratio effective/real radius&$\chi=0.328$\\
\hline
&Mass&$m_t=165[kg]$     \\
&Quality factor&$Q_t=5\cdot10^7$\\
&Resonator frequencies&$\frac{\omega_{t,k}}{2\pi}=1046,1051,1055,1059,1063,1068~[Hz]$\\
&Field in the capacity&$E=4.5\cdot10^7[V/m]$\\
Transducer&Capacities &$C_{t,k}=36.9,36.7,36.4,36.1,35.8,35.5~[nF]
$\\
&Transformer primary inductance&$L_p=1.3[H]$\\
&Transformer secondary inductance&$L_s=8\cdot 10^{-6}[H]$\\
&Transformer mutual inductance&$M=2.6\cdot10^{-3}[H]$\\
&Transformer resistance&$r_k\sim2.9\cdot10^{-4}[\Omega]$\\
 \hline
&Input inductance&$L_i=1.7\cdot10{-6}[H]$\\
SQUID&Washer inductance&$L_{SQ}=80\cdot10^{-12}[H]$\\
&Shunt resistance&$R_{sh}=4[\Omega]$\\
&Mutual inductance&$M_{SQ}=10[nH]$\\
 \hline
\end{tabular}
\end{center}
\caption{Parameter values for the filled sphere at the quantum
limit. If the mode ($j$) or transducer ($k$) indices is absent, its
means that the same value is used for all the indices range. The
frequency splitting in $\omega_{s,j}$ is extrapolated from the
miniGRAIL one. The capacities $C_{t,k}$ are chosen in order to tune
the electric frequencies on the resonator ones. The transformer
resistance is computed in order to get a electrical quality factor
$Q_e=5\cdot10^7$. The temperature and the SQUID effective
temperature are fixed to $20[mK]$. }\label{table 1}
\end{table}

\subsubsection{Single transducer results}

\begin{figure}
 \begin{center}
\includegraphics[width=0.8\textwidth]{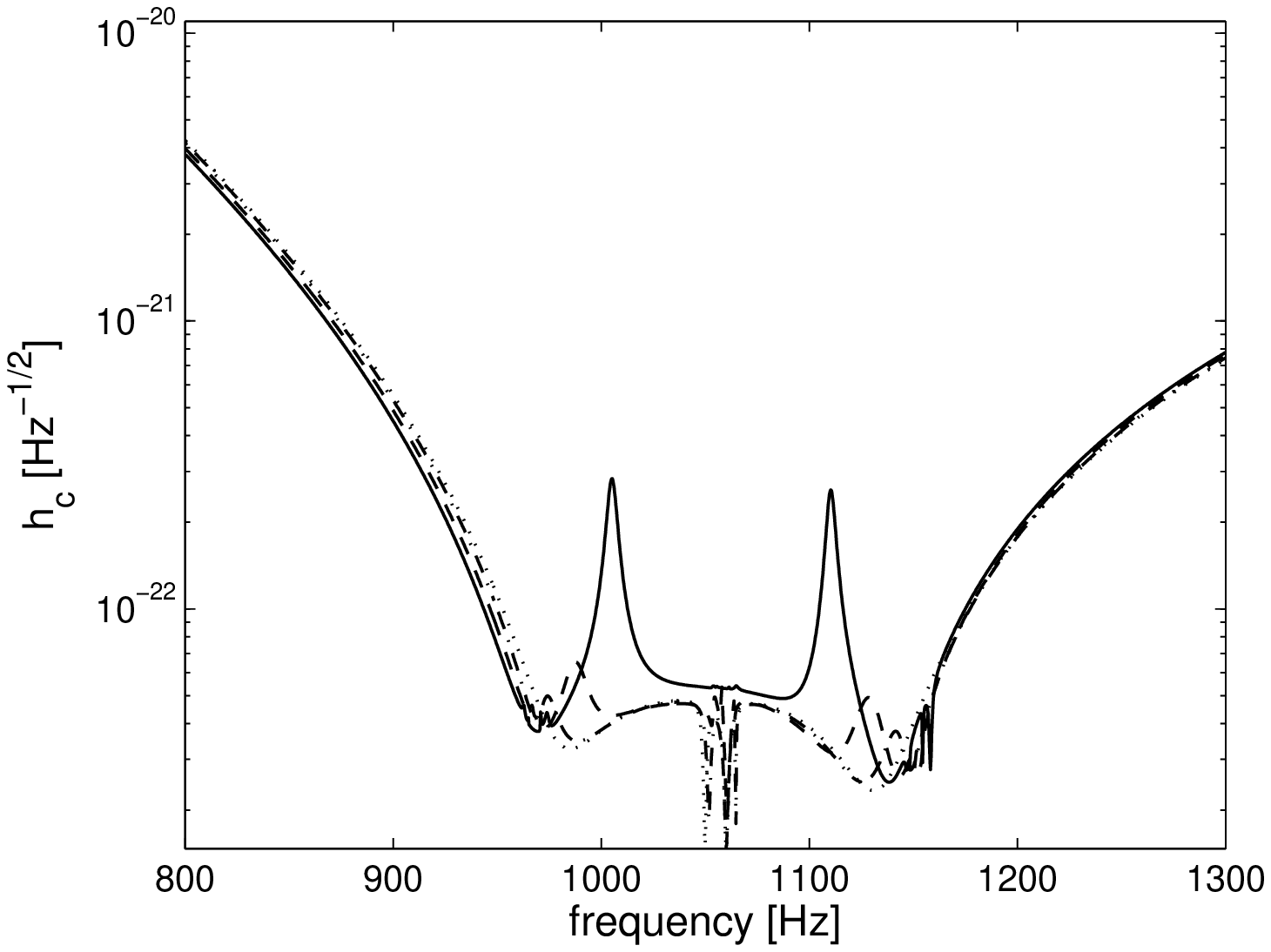}
\caption{\footnotesize Strain sensitivity for the single transducer
analysis. The purpose of this figure is to show the effect of the
noise due to the presence of other transducers. The dotted curve is
the sensitivity for a single transducer on a 1[m] radius bulk CuAl
sphere. Then, always performing the analysis of the output of one
transducer we add others transducers acting as noises sources: the
dashed-dotted curves stand for a sphere with 2 transducers on the 2
first TIGA locations, the dashed curves for a set of 4 transducers
on the 4 first TIGA locations, and the continuous curve is the
sensitivity for a transducer out of a 6 transducer TIGA
configuration. Note how the presences of further noises sources (=
the other transducers) deteriorate the sensitivity by adding horns.
 \label{fig SP2}}
 \end{center}
 \end{figure}

In the case of a single transducer analysis, we obtain $N$
sensitivity curves -- one for each transducer. As an example we
considerate a 1[m] radius bulk CuAl sphere with 6 transducers in
TIGA configuration. At the quantum limit, the typical strain
sensitivity of one of the transducers is plotted in Figure~\ref{fig
SP1}. Note the presence of horns on the both side of the resonance.
We can understand them as the contribution from the noises of the 5
others transducers. This explanation is confirmed by the sensitivity
curves obtained for the same sphere but with a different number of
transducers, see Figure~\ref{fig SP2}. As we increase the number of
transducers we also increase the number of noise sources, and the
sensitivity of a single transducer is worse than if the sphere was
equipped with a single transducer. Although this finding is not
surprising and can be related to existing
literature~\cite{Price:1987nz,Bassan:1988wq}, the details of this
feature are important in order to calibrate the transducers.

\begin{figure}
 \begin{center}
\includegraphics[width=0.8\textwidth]{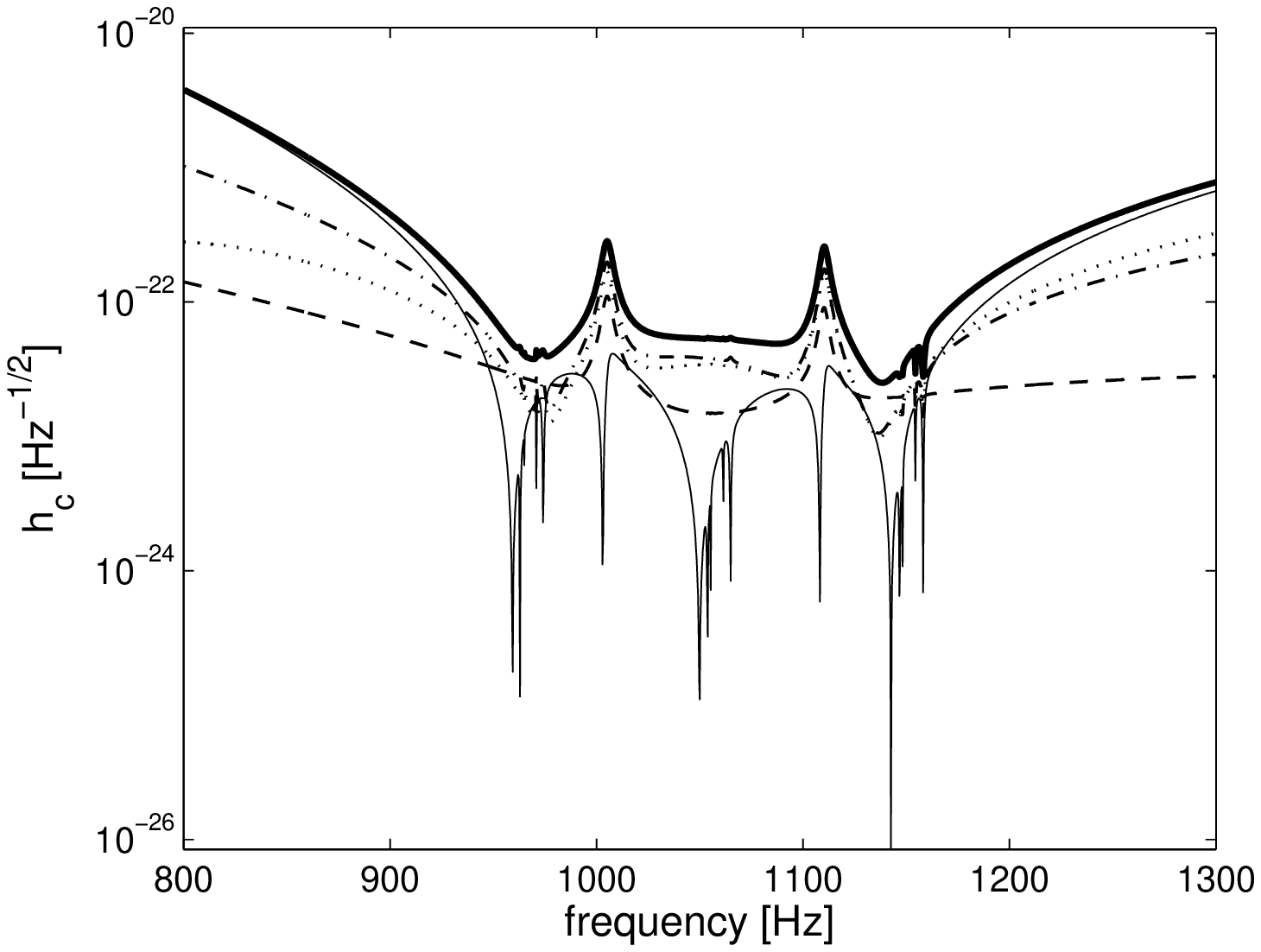}
\caption{\footnotesize
Strain sensitivity at the quantum limit for one of the 6 transducers placed into a TIGA configuration on a 1[m] radius bulk CuAl sphere. The curves are the relative contributions of the different noises: the thick line is the total sensitivity, the dashed curve is the mechanical thermal noise contribution (\ref{nm1},\ref{nm2}), the dashed-dotted line is the thermal electric noise contribution (\ref{ne1}), the dotted line the back-action contribution  (\ref{ban}) and the continuous curve the white noise (\ref{lnoise}).
 \label{fig SP1}}
 \end{center}
 \end{figure}

\subsubsection{Coherent analysis results}

\begin{figure}
 \begin{center}
\includegraphics[width=0.45\textwidth]{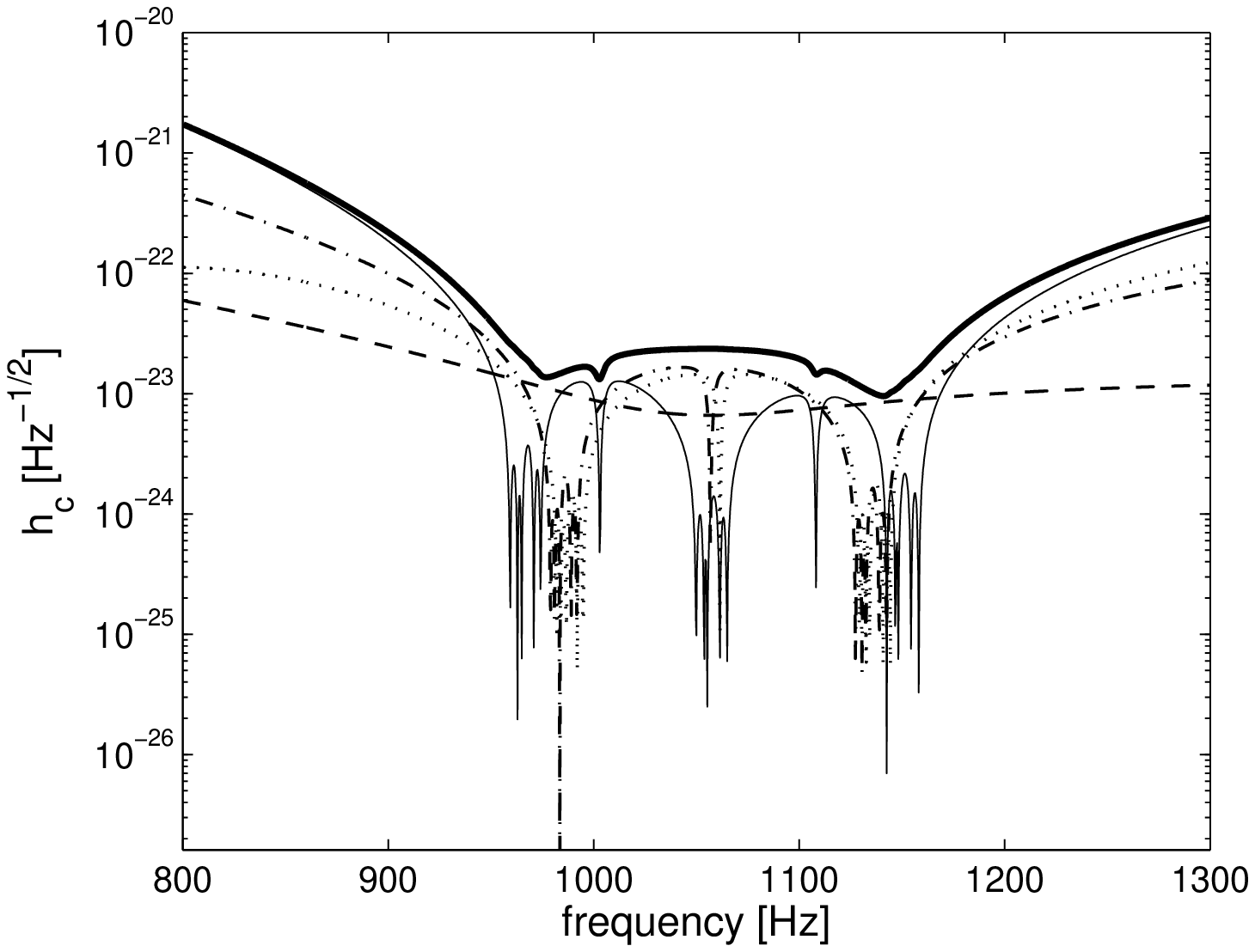}\includegraphics[width=0.45\textwidth]{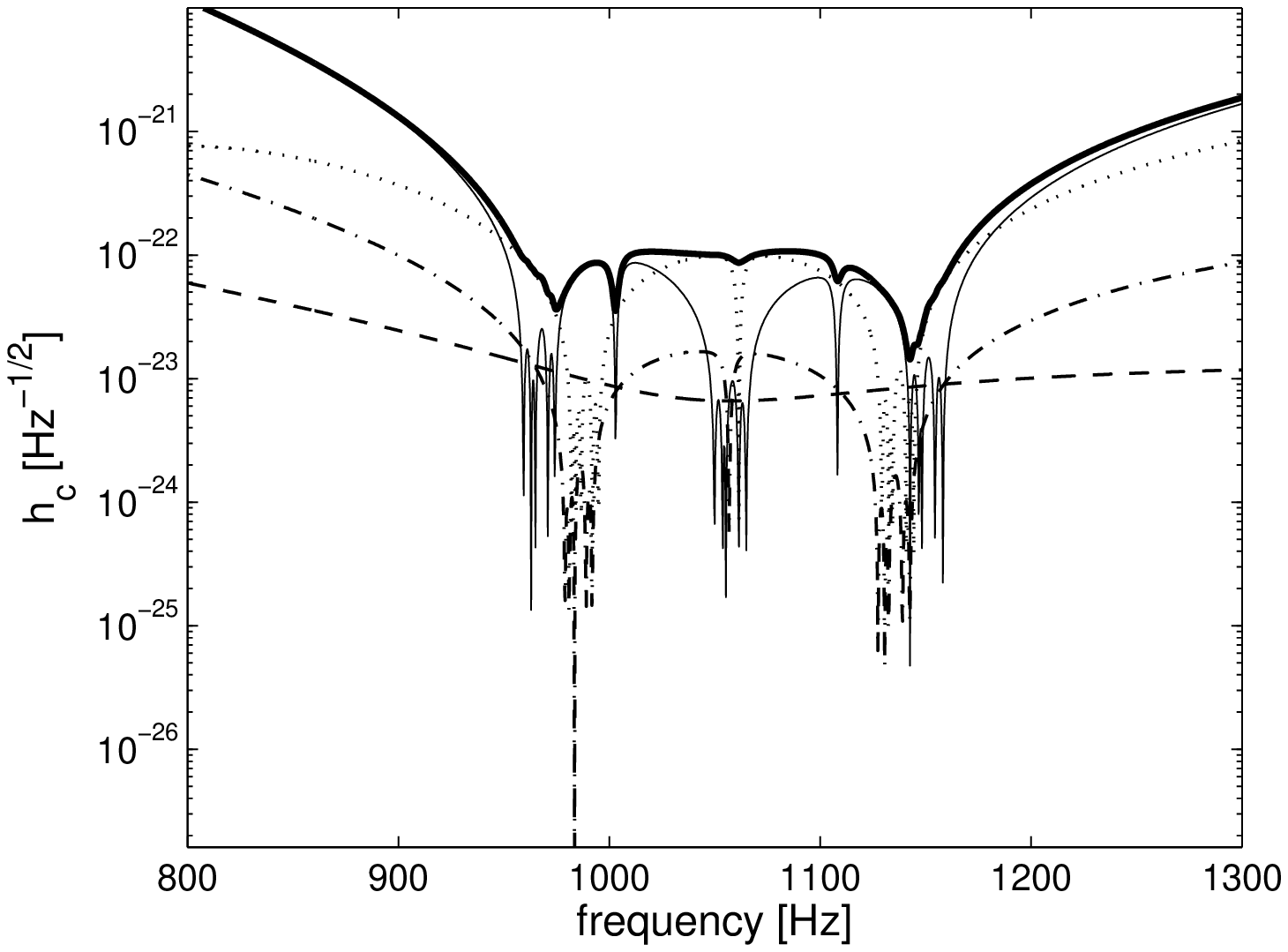}
\caption{\footnotesize Strain sensitivity at the quantum limit
(left) and for $N_{phonon}=50$ (right) for a coherent analysis of
the outputs of 6 transducers placed into a TIGA configuration on a
1[m] radius bulk CuAl sphere. The curves are the relative
contributions of the different noises: the thick line is the total
sensitivity, the dashed curve is the mechanical thermal noise
contribution (\ref{nm1},\ref{nm2}), the dashed-dotted line is the
thermal electric noise contribution (\ref{ne1}), the dotted line the
back-action contribution  (\ref{ban}) and the continuous curve the
white noise (\ref{lnoise}).   \label{fig SP3}}
 \end{center}
 \end{figure}

We now work out the coherent strain sensitivity for the same sphere
as in the previous case. For 6 transducers placed into a TIGA
configuration the sensitivity does not present horns as show in
Figure~\ref{fig SP3}. We also plot the sensibility corresponding to
the best present transducer ($N_{phonon}=50$).

\begin{figure}
 \begin{center}
\includegraphics[width=0.8\textwidth]{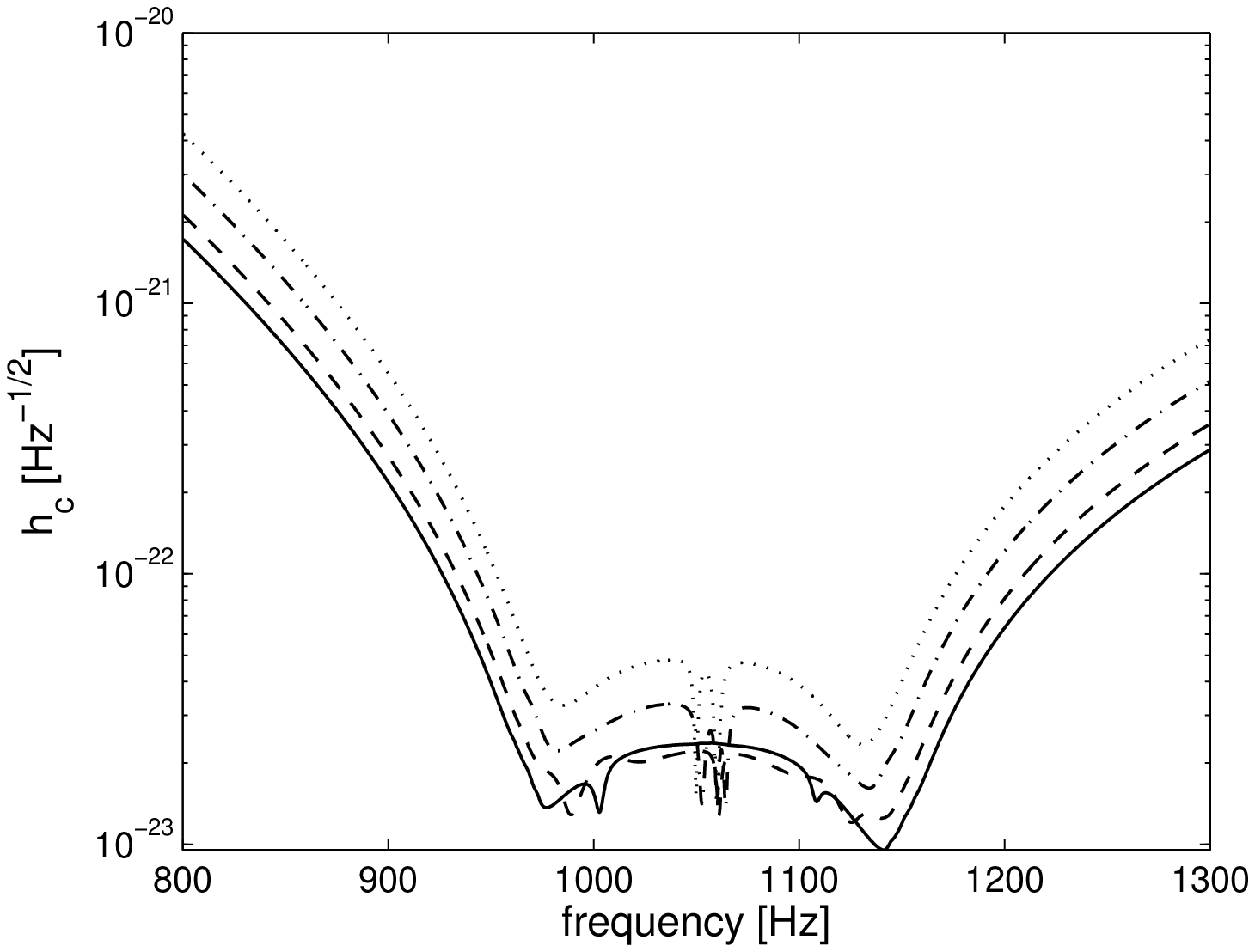}
\caption{\footnotesize Coherent strain sensitivity at the quantum limit of a transducer set placed on a 1[m] radius bulk CuAl sphere. The continuous curve is the sensitivity for a 6 transducers TIGA configuration, the dashed curves the one for a set of 4 transducers on the 4 first TIGA location the dashed-dotted curves the one for a set of 2 transducers on the 2 first TIGA location and the dotted curve is the sensitivity for a single transducer.  (Note that the dotted curve is the same as in Figure~\ref{fig SP2}) \label{fig SP4}}
 \end{center}
 \end{figure}

\modif{One can address the question of understanding how the
sensitivity is modified if we change the number of transducers. As
the sphere with a single transducer has the same pattern function as
a bar, we can expect that each new transducer partially complete the
coverage of the polarization and arrival directions until for $N=5$
we have a omnidirectional and omni-polarization detector. The
sensitivity will then be improved by the addition of each
transducer. Furthermore we will add the transducers in order to
complete a TIGA configuration and therefore, due to the particular
symmetry of this configuration, we expect the complete TIGA
configuration to be even more sensitive than the incomplete
5-transducers one. The simulations confirm this analysis as shown in
Figure~\ref{fig SP4}.}

\subsection{The hollow sphere}

\begin{figure}
 \begin{center}
\includegraphics[width=0.8\textwidth]{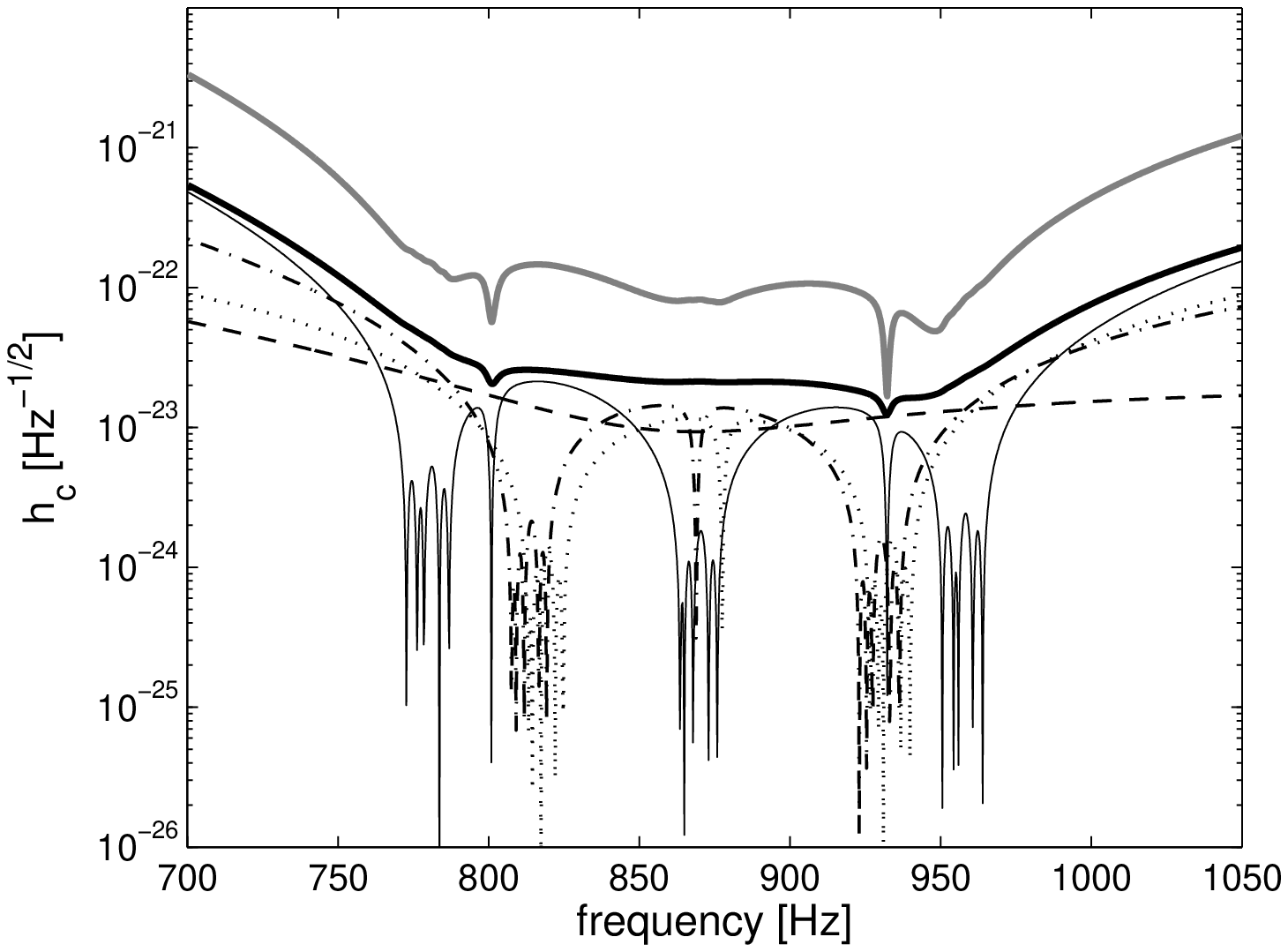} %[width=10cm]
\caption{\footnotesize Coherent strain sensitivity at the quantum limit for a set  of  6 transducers placed in a TIGA configuration on a 0.4[m]/1[m] internal/external radius CuAl sphere. The curves are the relative contributions of the different noises (see Figure~\ref{fig SP1} ). The gray curve is the sensibility corresponding to the best present transducer ($N_{phonon}=50$). \label{fig SC1}}
 \end{center}
 \end{figure}

All the above treatment has a nice, simple generalization to the
case of a hollow sphere  ~\cite{CocciaLobo}. While preserving all
the feature of a bulk sphere such as omnidirectionality, and the
capability to determine the source direction and wave polarization,
an hollow sphere has several interesting peculiar properties. Its
quadrupole frequencies are lower than  those of an equally massive
solid sphere, thus making the low-frequency range accessible to this
antenna with good sensitivity. Further, as shown in
~\cite{CocciaLobo}, for an appropriate ratio between the inner and
outer diameter, the cross section for the second quadrupole mode
equals that of the first, and one has the possibility of working
with a detector with the same (high) sensitivity at two frequencies.
The main differences with the bulk sphere are the numerical values
of the radial functions $\alpha_j$, that we first met in equation
(\ref{alpha}), and the one of the coupling to GW, namely $\chi$ (see
equation (\ref{chi})). We have also to recompute the
eigen-frequencies
(this can be done using the formulas from ref.~\cite{CocciaLobo}). \\
We have applied our model to a $1[m]$ external radius and $a=0.4[m]$
internal radius hollow sphere, with 6 transducers in the TIGA
configuration coupled to the first quadrupolar modes. The results of
our computation are displayed in the Table~\ref{table 2}. The
corresponding coherent strain sensitivity at the quantum limit is
show In Figure~\ref{fig SC1}.  Note that the frequency window is
shifted by about 200[Hz] with respect to the filled sphere and that
the sensitivity loss is only by about a factor 1.5.

\begin{table}[htdp]

\begin{center}
\begin{tabular}{|c|c|c|}

\hline
&Radius& $R_s=1[m]$\\
&Internal radius&$a=0.4[m]$\\
Sphere&Mass&$m_s=30974[kg]$     \\
&Modes frequencies&$\frac{\omega_{s,j}}{2\pi}=861,865,869,873,877~[Hz]$\\
&Radial eigen-function&$\alpha=-2.73$\\
&Ratio effective/real radius&$\chi=0.322$\\
\hline
&Mass&$m_t=155[kg]$     \\
&Resonator frequencies&$\frac{\omega_{t,k}}{2\pi}=859,863,867,871,875,879~[Hz]$\\
Transducer&Field in the capacity&$E=4.5\cdot10^7[V/m]$\\
&Capacities &$C_{t,k}= 55,54,54,53,53,52~[nF]
$\\
&Transformer resistance&$r_k\sim2.4\cdot10^{-4}[\Omega]$\\
 \hline

\end{tabular}
\end{center}
\caption{Parameters value for the hollow sphere with $a=0.4[m]$
internal radius at the quantum limit. The transducer being build on
the same model as for the filled sphere we put in this table only
the quantities which differ from the filled case. The temperature
and the SQUID effective temperature are fixed to $20[mK]$. The other
values can be found into the Table~\ref{table 1}.\label{table 2}}
\end{table}

 \subsubsection{Effect of the thickness of the hollow sphere}

\begin{figure}
 \begin{center}
\includegraphics[width=0.8\textwidth]{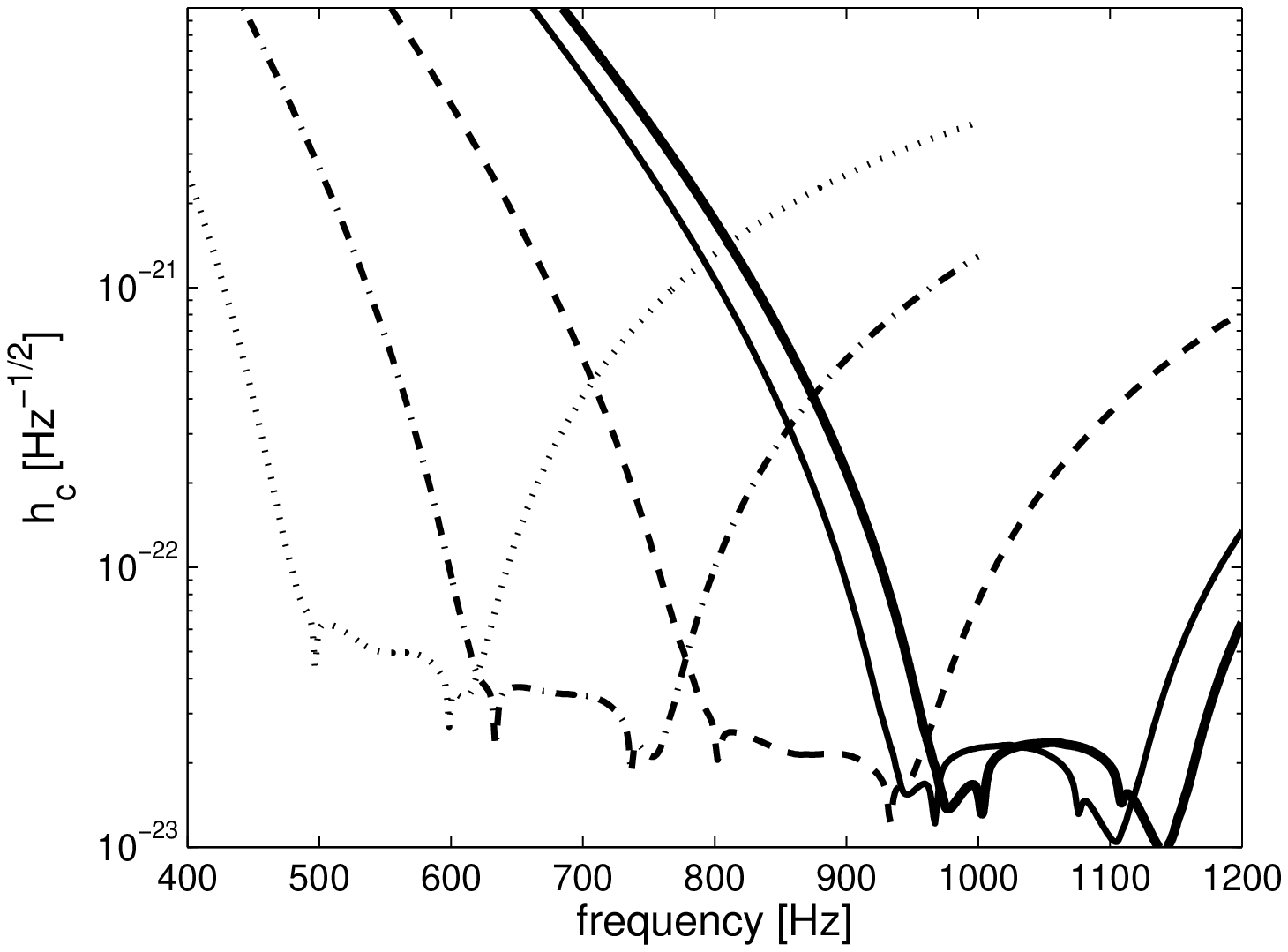} %[width=10cm]
\caption{\footnotesize Coherent strain sensitivities at the quantum limit for a  TIGA configuration on a 1[m] external radius hollow CuAl sphere for different values of the internal radius $a$. The dotted curve is for $a=0.8[m]$ (ie. a shell of thickness $0.2[m]$) the successive curves are for $a$ decreasing by step of $0.2[m]$ until the black thick line which stand for the filled sphere.   \label{fig SCa}}
 \end{center}
 \end{figure}

We address now the question of the influence of the thickness of the
sphere on its sensitivity. For a fixed external radius a thinner
sphere is less massive, and therefore less sensitive. There is also
a shift in the resonances to lower frequencies when the sphere is
thinner, see Figure~\ref{fig SCa}.

\subsubsection{Enlarging the bandwidth by monitoring the second quadrupolar multiplet}

We now compare different designs of multi-modal resonant detectors
based on a hollow sphere. We will always work on the base of our
0.4/1[m] hollow sphere described above and we compare different
transducers designs.

Having a description of the hollow sphere with 6 transducers
monitoring the first quadrupolar multiplet we can now simulate a
multi-modal detector. As discussed in  Section~\ref{S coupling}, we
have to take into account the presence of other sphere modes. In
Table~\ref{table 3}, we list the properties of the modes we include
in the simulation of our multi-modal models. Each modes multiplet
$(n,\ell)$ contain $2\ell+1$ modes and therefore we take into
account a total of $J=30$ spheroidal modes in our simulations. Only
10 of them are coupled to GW.
\begin{table}[htdp]

\begin{center}
\begin{tabular}{|c|c|c|c|c|}

\hline
$n,\ell$&$f_{n,l}$&$\alpha_{n,l}(R_S)$&$\alpha_{n,l}(a)$&$\chi_{n,l}$\\
 \hline
 (1,2)&869[Hz]&-2.73&-2.85&0.319\\
 (1,3)&1421[Hz]&-0.488&17.1&0\\
 (1,1)&1558[Hz]&-1.41&-4.02&0\\
 (1,0)&1758[Hz]&-1.29&-1.06&0\\
 (1,4)&1937[Hz]&1.51&21.1&0\\
 (2,2)&1970[Hz]&0.276&-3.27&-0.148\\
 other&$>$2.4[kHz]&&&\\
 \hline

\end{tabular}
\end{center}
\caption{Frequency, radial eigen-function (at the outer and inner
surface) and coupling to GW for the first modes of a hollow sphere
with external radius $R_S=1[m]$ and internal radius $a=0.4[m]$. All
these quantities are computed from formulas found
in~\cite{CocciaLobo}.\label{table 3}}
\end{table}

For a hollow sphere, we have the choice of placing the transducers
on the outer or on the inner surface (this choice may be limited by
the experimental difficulty of placing transducers inside the
sphere). Looking only at the quadrupolar modes, and because the
radial displacement of the second mode is maximum at the inner side
of the sphere ~\cite{CocciaLobo}, one could think of improving the
sensitivity by mounting the classical capacitive transducers
considered here inside the sphere. However, as show in
Table~\ref{table 3}, the higher multipolar modes have very large
$\alpha$'s on the inner surface. These modes will
couple to the transducers main resonances and their thermal
noise will eventually limit the detector sensitivity. This effect is
analysed in detail below.

A first approach to design a multi-modal spherical detector is to
add a second set of 6 transducers in TIGA configuration coupled to
the second quadrupolar multiplet.
The characteristic of the second set of transducer are listed in Table~\ref{table 4}. The parameters are optimized to get the best strain sensitivity and the largest bandwidth following the detailed analysis for a bulk sphere reported in \cite{Gottardi,GottardiPRD}. We notice that the optimal mass of the resonator coupled to the second quadrupolar modes is smaller that the first resonator. This is due to the fact that the effective mass of the higher quadrupolar modes  is smaller and the higher frequencies of the modes requires a smaller transducer mass to have the same electro-mechanical coupling.    
 In our simulation we can choose to place the transducers inside or outside the sphere. On Figure~\ref{fig
SCdouble}  we show the difference in the strain sensitivity  for a
hollow sphere with 12 transducers when the thermal noise contribution
of all the modes between the first and second quadrupolar multiplet is included.  Both the
configurations with the second set of
transducers placed outside and inside the sphere are considered. The
thermal noise of the modes not coupled to GW  reduces slightly the
bandwidth in the former 
case, but has dramatic effect on the sensitivity in the latter case.

\begin{figure}
 \begin{center}
\includegraphics[width=0.5\textwidth]{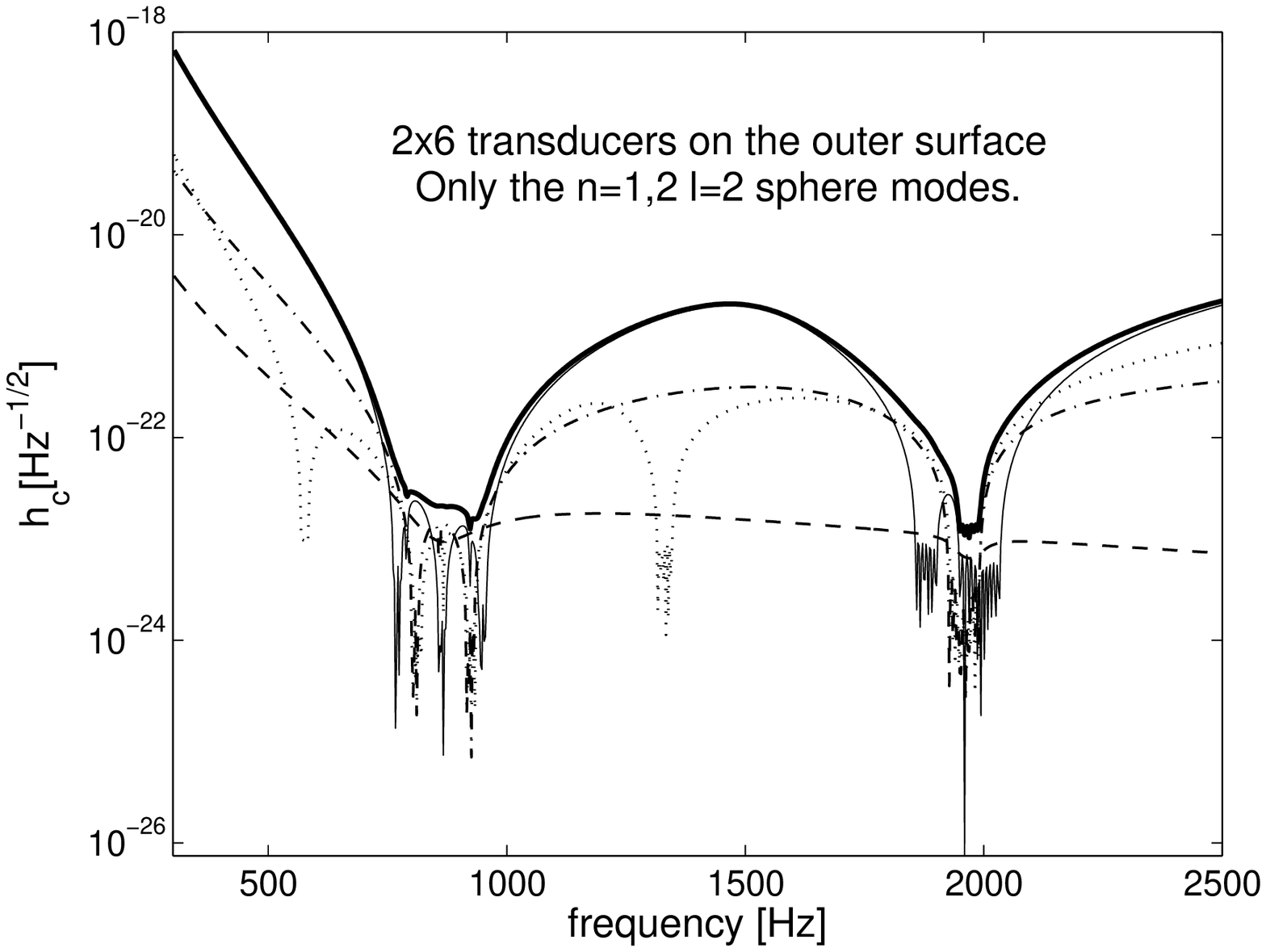}~\includegraphics[width=0.5\textwidth]{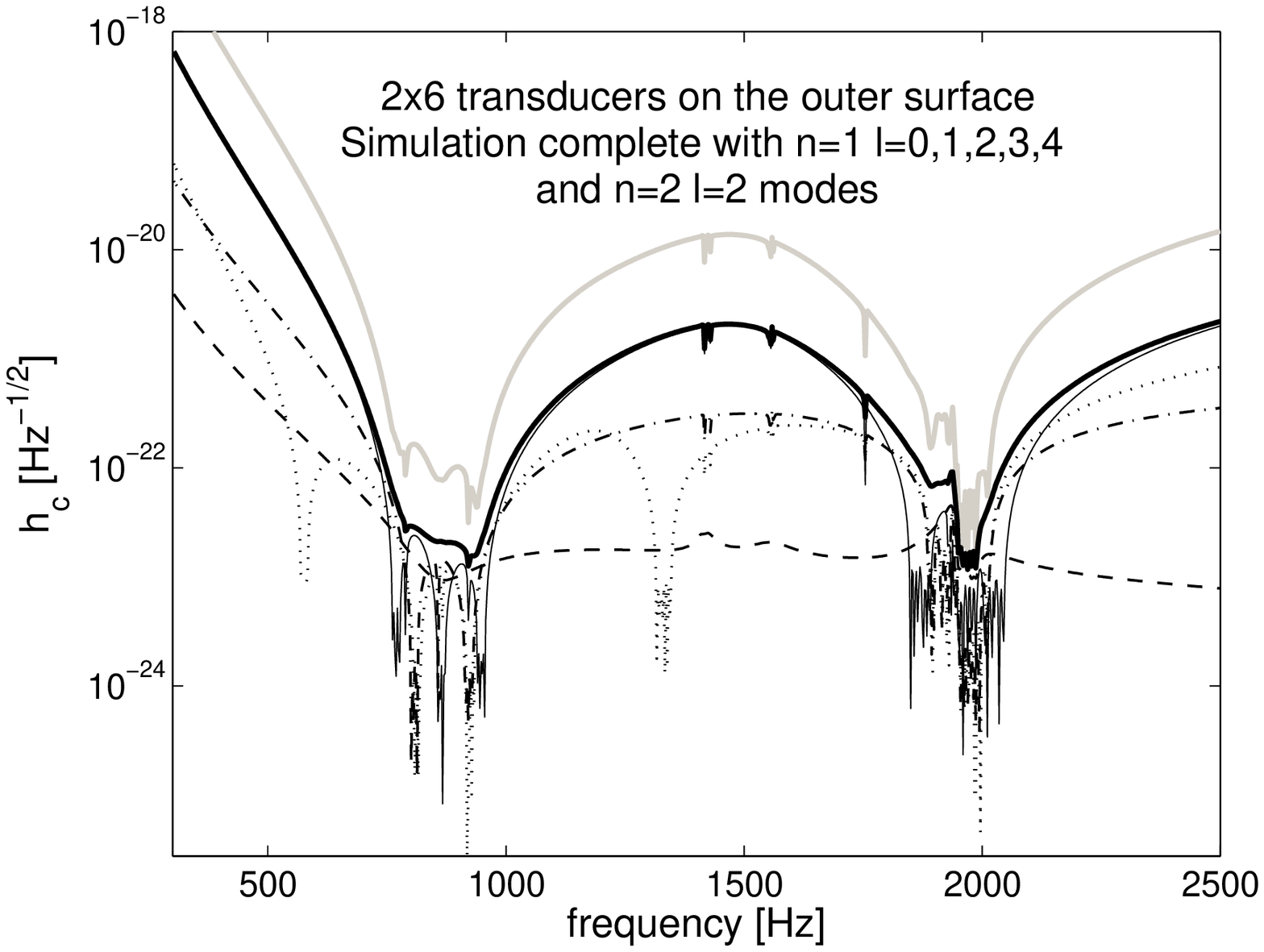} \\
\includegraphics[width=0.5\textwidth]{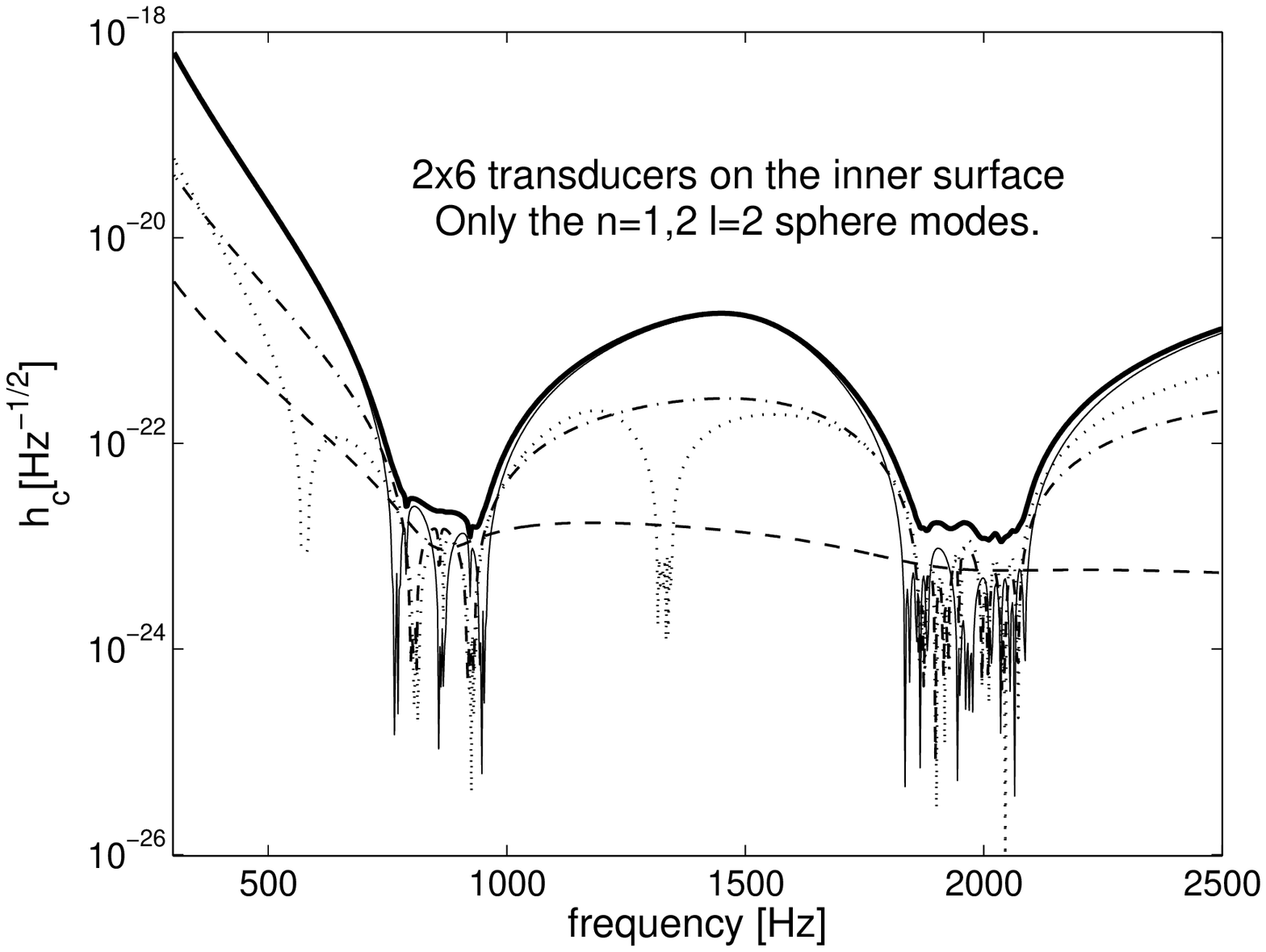}~\includegraphics[width=0.5\textwidth]{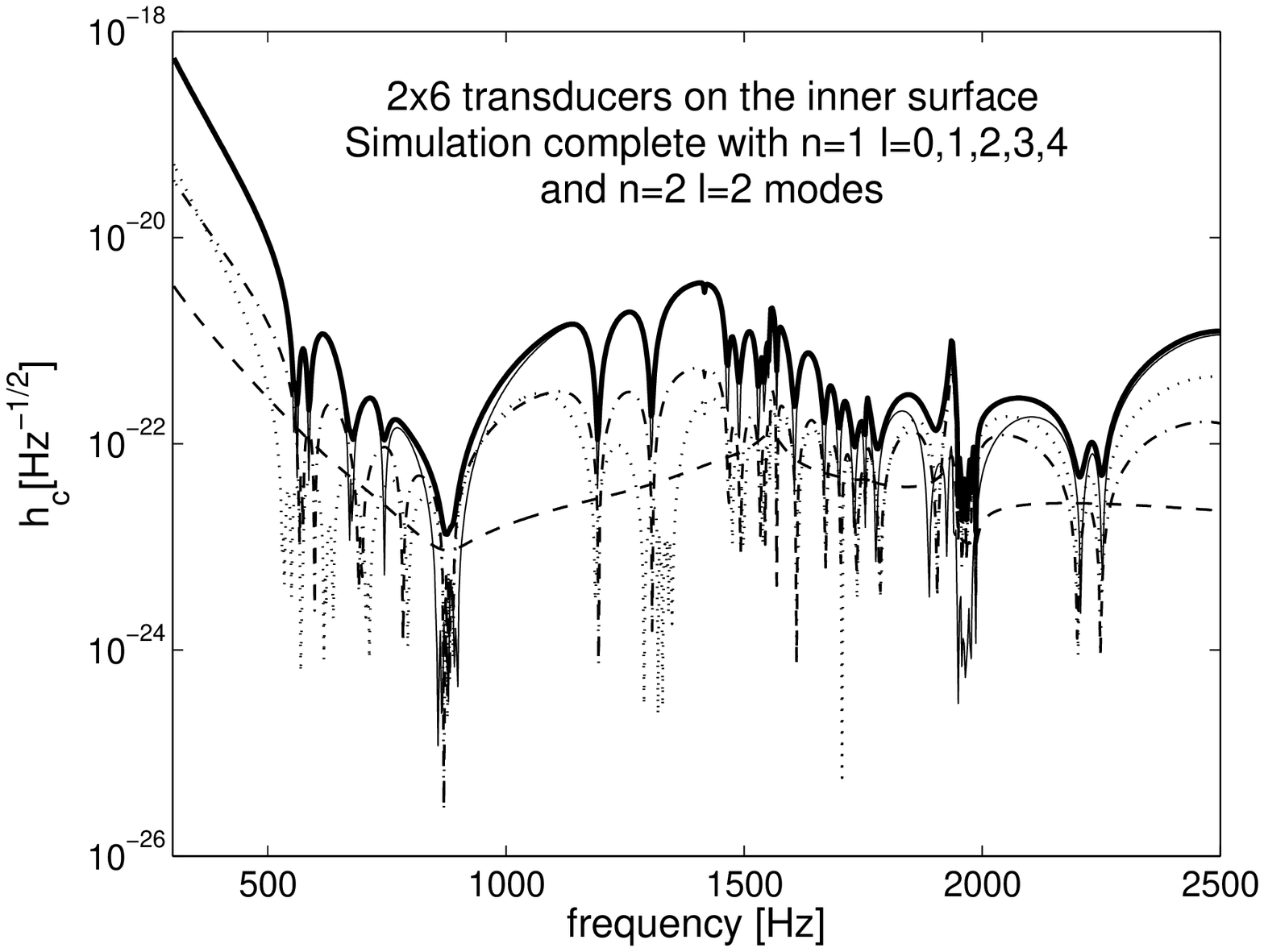}
\caption{\footnotesize  Coherent strain sensitivity at the quantum limit for two sets  of  6 transducers placed in TIGA configuration on a 0.4[m]/1[m] internal/external radius CuAl sphere. Each transducer set is coupled to the first and second quadrupolar modes, respectively.  The curves show the relative contributions of the different noises: the black thick line is the total sensitivity, the dashed curve is the total mechanical thermal noise contribution (\ref{nm1},\ref{nm2}), the dashed-dotted line is the thermal electric noise contribution (\ref{ne1}), the dotted line the back-action contribution  (\ref{ban}) and the continuous curve the white noise (\ref{lnoise}).  The left panels are the result of the simulations with only the quadrupolar sphere modes, in the right panel we have add the other modes with frequency smaller than $2.5[kHz]$ The upper plots are for transducers on the outer surface and the lower ones are for transducer on the inner surface. On the upper right plot the gray curve is the sensitivity for $N_{phonon}=50$\label{fig SCdouble}}
 \end{center}
 \end{figure}

 \begin{table}[htdp]

\begin{center}
\begin{tabular}{|c|c|c|}

\hline

Mass&$m_{t,k=7..12}\sim\frac{1}{5}m_{t,k=1..6}=31[kg]$     \\
Resonator frequencies&$\frac{\omega_{t,k}}{2\pi}=1947,1956,1965,1974,1983,1992~[Hz]$\\
Field in the capacity&$E=4.5\cdot10^7[V/m]$\\
Capacities &$C_{t,k}\sim10~[nF]
$\\
Transformer resistance&$r_k\sim5.4\cdot10^{-4}[\Omega]$\\
 \hline

\end{tabular}
\end{center}
\caption{Parameter values for the second set of transducers. For
comments, see Table~\ref{table 1} and for the sphere parameters see
Table~\ref{table 2} and ~\ref{table 3}  .\label{table 4}}
\end{table}

From Figure~\ref{fig
SCdouble} one can conclude that placing the transducers inside the
sphere will not lead to an improvement of the sensitivity. We
observe that a 12 transducers in 2 TIGA configuration on the outer
surface of a 0.4/1[m] hollow sphere can lead to a sensitivity curve
with 2 windows below $3\cdot10^{-23}[Hz^{-1/2}]$ with bandwidth of
175 and 52[Hz].

Having in mind that the presence of 12 transducers can represent an
experimental problem (even on the outer surface of the sphere), we
now consider the possibility of reducing the number of transducers
by using the double-mode transducer described in
Section~\ref{SQUID2}. We then perform simulations of a multi-modal
spherical GW detector with only 6 double-mode transducers in a TIGA
configuration.

\begin{figure}
 \begin{center}
\includegraphics[width=0.8\textwidth]{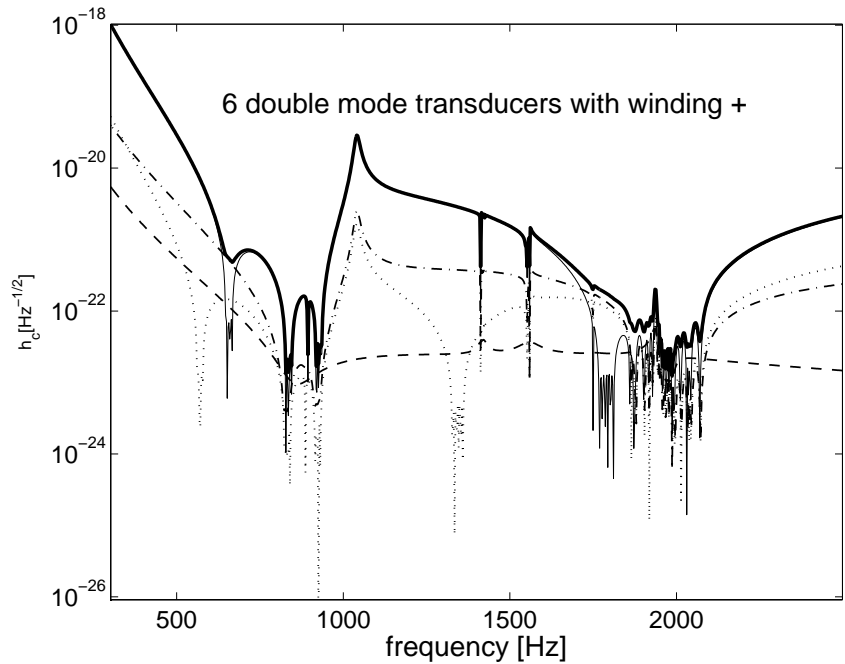}
\caption{\footnotesize  Coherent strain sensitivity at the quantum limit for a set of 6 double-mode transducers placed in TIGA configuration on a 0.4[m]/1[m] internal/external radius CuAl sphere.  The curves show the relative contributions of the different noises: the black thick line is the total sensitivity, the dashed curve is the total mechanical thermal noise contribution (\ref{nm1},\ref{nm2}), the dashed-dotted line is the thermal electric noise contribution (\ref{ne1}), the dotted line the back-action contribution  (\ref{ban}) and the continuous curve the white noise (\ref{lnoise}). All the spheroidal modes with frequency smaller than 2.5[kHz] are included in the simulation.\label{pm}}
 \end{center}
 \end{figure}

In figure~\ref{pm} we show the sensitivity of a hollow sphere equipped
with 6 double-mode transducers. The transducers parameters, reported in Table~\ref{table 5}, are chosen to get the
best sensitivity and the largest bandwidth at the two principal
quadrupolar modes multiplets around 800 and 1900 Hz.  The choice of the parameters in
Table~\ref{table 5} is the result of an
iterative process based on the detailed optimization procedure
described in \cite{Gottardi,GottardiPRD} for a bulk sphere. A more detailed technical analysis is
necessary to fully optimized the detector described here. Such an analysis is beyond the scope of this paper. 
 \begin{table}[htdp]
\begin{center}
\begin{tabular}{|c|c|}
\hline
Mass&$m_{t,k=7..12}\sim \frac{1}{5}m_{t,k=1..6}=31[kg]$      \\
Quality factor&$Q_{t1}=Q_{t2}=5\cdot10^7$\\
Resonator frequencies&$\frac{\omega_{t1}}{2\pi}=859,863,867,871,875,879~[Hz]$\\
&$\frac{\omega_{t2}}{2\pi}=1947,1956,1965,1974,1983,1992~[Hz]$\\
Field in the capacity&$E_1=E_2=4.5\cdot10^7[V/m]$\\
Capacities &$C_{1,k}\sim 35~[nF]
$\\
&$C_{2,k}\sim 10~[nF]$\\
Transformer primary inductance&$L_{1,p}=L_{2,p}=1.3[H]$\\
Transformer secondary inductance&$L_{1,s}=L_{2,s}=8\cdot 10^{-6}[H]$\\
Transformer mutual inductance&$M_1=M_2=2.6\cdot10^{-3}[H]$\\
Transformer resistance&$r_{1,k}\sim2.4\cdot10^{-4}[\Omega]$\\
&$r_{2,k}\sim5.4\cdot10^{-4}[\Omega]$\\
\hline
Input inductance&$L_i=1.6\cdot10{-6}[H]$\\
Washer inductance&$L_{SQ}=80\cdot10^{-12}[H]$\\
Shunt resistance&$R_{sh}=4[\Omega]$\\
Mutual inductance&$M_{SQ}=10[nH]$\\
 \hline
\end{tabular}
\end{center}
\caption{Parameters value for the set of double modes transducers.
For comments see Table~\ref{table 1} and for the sphere parameters
see Table~\ref{table 2} and ~\ref{table 3}.\label{table 5}}
\end{table}

The sensitivity around the two most sensitive spheroidal modes is
comparable with the one obtained with a 12 single-mode transducers
configuration. Thanks to the use of a single SQUID to amplify the
signal from both the spheroidal modes families, the resulting
back-action noise contribution of the SQUID amplifiers is reduce in
the double-transducer read-out scheme. This leads to a larger
bandwidth at the second spheroidal mode frequencies. Between the two
modes families the sensitivity is reduced due to a low mechanical
coupling between the resonators and the sphere modes, and is limited
by the additive SQUID white noise. A strain sensitive level of about
$10^{-22} \; Hz^{-1/2}$ and a total bandwidth of about 600 Hz could
be achieved in this way around 800 and 2000 Hz. At the sensitivity
of the currently best performing resonant bar antenna \cite{Auriga},
a multimodal hollow spherical detector could reach a bandwidth of
about 1kHz.

\subsection{Concluding remarks\label{c}}

We developed a mathematical framework, which can describe the
indirect coupling of the transducers through the modes of the
sphere. We have shown how to build models giving the strain
sensitivity of a GW spherical detector and presented different
configurations of a multi-mode spherical resonant detector. The
effect of the thermal noise from higher frequency modes of the
sphere, not coupled to GW, are also included in the model. We fully
analyzed the strain sensitivity of an hollow CuAl sphere of
0.4[m]/1[m] internal/external radius. Such an antenna, equipped with
12 `standard' transducers or 6 double-mode transducers, when
optimally tuned, may display a sensitivity curve, at the quantum
limit, below $3\cdot10^{-23}[Hz^{-1/2}]$ with a total bandwidth up
to 600 Hz around the first two quadrupolar multiplets at about 800
and 2000 Hz. At the sensitivity of the currently best performing
resonant bar antenna \cite{Auriga}, a multimodal hollow spherical
detector could reach a bandwidth of about 1kHz. Such a  detector
could have peak sensitivity comparable with the first generation of
interferometers, but would be also able to determine the GW arrival
direction. With an appropriate resizing of the sphere, the second
quadrupolar modes can be shifted to the frequency region where
existing small spherical detectors are sensitive. This would open
the possibility of coincidence search between several spherical
detectors and the DUAL detector \cite{Bonaldi06}, building in this
way  the base for a powerful omnidirectional gravitational wave
observatory.

\section*{Acknowledgements}
We wish to acknowledge M. Maggiore, S. Foffa, M.A. Gasparini, C. Caprini, A. Malaspinas and L. Bonacina for many useful discussions. \\
The work of J.E. is partially supported by the Swiss National Funds.
The work of F.D. is supported by the Swiss National Funds and by the
National Science Foundation under Grant No. PHY99-07949. The work of
L.G.  was partially supported by the  Integrated  Large
Infrastructures  for Astroparticle Science (ILIAS) of  the Sixth
Framework Program of the European Community when the author was
employed  at Leiden University.

\end{document}